\documentclass[aps,prd,onecolumn,amsmath,showpacs,superscriptaddress,nofootinbib,nopreprintnumbers,tightenlines,notitlepage]{revtex4-1}

\usepackage{verbatim}
\usepackage[T1]{fontenc}
\usepackage[utf8]{inputenc}
\usepackage[american]{babel}
\usepackage{epsfig}
\usepackage{graphicx}

\usepackage{hyperref}

\usepackage{booktabs}
\usepackage{multirow}
\usepackage{dcolumn}
\usepackage{amsmath}
\usepackage{mathtools}
\usepackage{amsfonts}
\usepackage{amssymb}
\usepackage{epstopdf}
\usepackage{bm}
\usepackage{siunitx}
\usepackage{braket}
\usepackage{enumitem}
\usepackage{soul}
\usepackage[table]{xcolor}
\usepackage{color}
\usepackage{transparent}

\usepackage{enumitem}
\usepackage{graphicx}
\usepackage[font=small]{caption}
\usepackage{subcaption}
\usepackage{pifont}

\definecolor{navyblue}{rgb}{0.0, 0.0, 0.5}
\definecolor{royalblue}{rgb}{0.25, 0.41, 0.88}
\definecolor{cadmiumgreen}{rgb}{0.0, 0.42, 0.24}
\definecolor{blue-violet}{rgb}{0.54, 0.17, 0.89}
\definecolor{darkviolet}{rgb}{0.58, 0.0, 0.83}
\definecolor{teal(colorwheel)}{rgb}{1.0, 0.5, 0.0}

\usepackage{hyperref}
\hypersetup{
    colorlinks=true, 
    linkcolor=royalblue, 
    citecolor=magenta}

\newcommand\ee{\end{equation}}
\newcommand\be{\begin{equation}}
\newcommand\eea{\end{eqnarray}}
\newcommand\bea{\begin{eqnarray}}

\newcommand\ie{{\it i.e.}~}

\usepackage{booktabs}
\usepackage{multirow}
\usepackage{dcolumn}
\usepackage{colortbl}

\definecolor{magenta(process)}{rgb}{1.0, 0.0, 0.56}

\definecolor{darkspringgreen}{rgb}{0.09, 0.45, 0.27}

\definecolor{royalblue(web)}{rgb}{0.25, 0.41, 0.88}
\newcommand{\nq}[1]{%
	\begin{tabular}{@{}c@{}}\strut#1\strut\end{tabular}%
}

\begin{document}

\title{Health checkup test of the standard cosmological model in view of recent Cosmic Microwave Background Anisotropies experiments.}

\author{Eleonora Di Valentino}
\email{e.divalentino@sheffield.ac.uk}
\affiliation{School of Mathematics and Statistics, University of Sheffield, Hounsfield Road, Sheffield S3 7RH, United Kingdom}

\author{William Giar\`e}
\email{william.giare@uniroma1.it}
\affiliation{Galileo Galileo Institute for theoretical physics, Centro Nazionale INFN di Studi Avanzati, \\ Largo Enrico Fermi 2,  I-50125, Firenze, Italy}
\affiliation{INFN Sezione di Roma, P.le A. Moro 2, I-00185, Roma, Italy}

\author{Alessandro Melchiorri}
\email{alessandro.melchiorri@roma1.infn.it}
\affiliation{Physics Department and INFN, Universit\`a di Roma ``La Sapienza'', Ple Aldo Moro 2, 00185, Rome, Italy}

\author{Joseph Silk}
\affiliation{Institut d'Astrophysique de Paris (UMR7095: CNRS \& UPMC- Sorbonne Universities), F-75014, Paris, France}
\affiliation{Department of Physics and Astronomy, The Johns Hopkins University Homewood Campus, Baltimore, MD 21218, USA}
\affiliation{BIPAC, Department of Physics, University of Oxford, Keble Road, Oxford OX1 3RH, UK}

\date{\today}


\begin{abstract}

We present an updated data-analysis comparison of the most recent observations of the Cosmic Microwave Background temperature anisotropies and polarization angular power spectra released by four different experiments: the Planck satellite on one side, and the Atacama Cosmology Telescope (ACTPol) and the South Pole Telescope (SPT-3G), combined with the WMAP satellite 9-years observation data in order to be "Planck-independent" on the other side. We investigate in a systematic way 8 extended cosmological models that differ from the baseline $\Lambda$CDM case by the inclusion of many different combinations of additional degrees of freedom, with the aim of finding a viable minimal extended model that can bring all the CMB experiments in agreement. Our analysis provides several hints for anomalies in the CMB angular power spectra in tension with the standard cosmological model that persist even in these multi-parameter spaces. This indicates that either significant unaccounted-for systematics in the CMB data are producing biased results or that $\Lambda$CDM is an incorrect/incomplete description of Nature. We conclude that only future independent high-precision CMB temperature and polarization measurements could provide a definitive answer.

\end{abstract}

\maketitle

\section{Introduction}
\label{sec.introduction}

The measurements of Cosmic Microwave Background (CMB) temperature anisotropies and polarization angular power spectra provided by the Planck satellite~\cite{Planck:2019nip,Planck:2018nkj}, while broadly in agreement with the expectations of the standard $\Lambda$CDM model of cosmology~\cite{Planck:2018vyg}, show intriguing anomalies that clearly deserve further investigation. In particular, the Planck data show a higher lensing amplitude at about $2.8$ standard deviations~\cite{DiValentino:2019dzu,DiValentino:2020hov}. Since more lensing is expected with more Cold Dark Matter (CDM), this immediately recasts a preference for a closed Universe that also helps in explaining some large-scale anomalies in the data, like the deficit of amplitude in the quadrupole and octupole modes. Consequently, the final indication for a closed Universe reaches the level of $3.4$ standard deviations~\cite{DiValentino:2019qzk,Handley:2019tkm}, even higher than the lensing amplitude anomaly. This is clearly an unexpected result since there are several other complementary astrophysical observables, like Baryon Acoustic Oscillation (BAO) measurements~\cite{Beutler:2011hx,Ross:2014qpa,Alam:2016hwk}, that are in tension with a closed Universe when combined with the Planck data~\cite{DiValentino:2019qzk}. In addition, the widely known tension between the value of the Hubble constant measured by the SH0ES collaboration using luminosity distances of Type Ia supernovae~\cite{Riess:2021jrx} ($H_0=73\pm1$ Km/s/Mpc) and the value inferred by the Planck satellite from the CMB observations~\cite{Planck:2018vyg} ($H_0=67.4\pm0.5$ Km/s/Mpc) recently crossed the symbolic limit of 5 standard deviations (see also Ref.~\cite{Riess:2022mme} where the disagreement reached $5.3\sigma$), basically ruling out the possibility of a statistical fluke. The intriguing characteristic is that none of the late time universe measurements, that agree very well with the SH0ES result, give a value below the early universe measurements, that agree very well with Planck, and vice-versa. Among them we can list the late universe results based on the SNIa calibrated with the Tip of the Red Giant Branch~\cite{Freedman:2021ahq,Anand:2021sum,Jones:2022mvo,Dhawan:2022yws},  the Surface Brightness Fluctuation measurements~\cite{Blakeslee:2021rqi,Khetan:2020hmh,Garnavich:2022hef}, the Type II Supernovae~\cite{deJaeger:2020zpb,deJaeger:2022lit}, the Megamaser Cosmology Project~\cite{Pesce:2020xfe}, the strongly lensed quasars~\cite{Wong:2019kwg,Liao:2020zko,Qi:2020rmm,Yang:2020eoh,Birrer:2020tax,Denzel:2020zuq}. While in the early universe we have the alternative CMB data released by the Atacama Cosmology Telescope (ACT)~\cite{ACT:2020frw,ACT:2020gnv} and South Pole Telescope (SPT)~\cite{SPT-3G:2014dbx,SPT-3G:2021eoc}, and large scale structure measurements~\cite{eBOSS:2020yzd,DAmico:2019fhj,Philcox:2020vvt,Philcox:2020xbv,Farren:2021grl}, all of them preferring a lower $H_0$ value and assuming the standard $\Lambda$CDM model. Since inferring the value of the present day expansion rate from observations of the early Universe (\ie the CMB) necessarily requires a cosmological model, this tension seriously questions the validity of the baseline $\Lambda$CDM scenario, reveling either the presence of important observational systematics in the data or the need for new physics (see, \textit{e.g.}, Ref.~\cite{DiValentino:2020zio,Jedamzik:2020zmd,CANTATA:2021ktz,DiValentino:2021izs,Perivolaropoulos:2021jda,Schoneberg:2021qvd,Abdalla:2022yfr} for recent reviews). Also, we shouldn't forget the $S_{8} \equiv \sigma_{8} \sqrt{\Omega_{m} / 0.3} $ tension between the value estimated by Planck and the weak lensing measurements KiDS-1000~\cite{Heymans:2020gsg} or DES-y3~\cite{DES:2021vln}, assuming a $\Lambda$CDM model, at $2-3\sigma$ level. Therefore, precise measurements of the cosmological parameters beyond Planck are certainly needed to shed light on the nature of these tensions and anomalies as well as to independently test the theoretical assumptions underlying the standard paradigm of modern cosmology. 

Thanks to the new measurements of CMB temperature anisotropies and polarization recently released by ACT~\cite{ACT:2020frw,ACT:2020gnv} and SPT~\cite{SPT-3G:2014dbx,SPT-3G:2021eoc}, this is now possible.  Indeed, when combined with the WMAP 9-year observation data, these experiments produce cosmological bounds that, while typically less constraining than those obtained by the Planck collaboration~\cite{ACT:2020gnv,SPT-3G:2021wgf,SPT:2017sjt}, may be enough tight for accurate independent tests of the Planck results~\cite{Handley:2020hdp}. 

Interestingly, the combination of the Atacama Cosmology Telescope and WMAP data
fully confirms the inflationary prediction for a spatially flat Universe~\cite{ACT:2020gnv}, possibly leading weight to the hypothesis of a statistical fluctuation or an undetected systematic as a solution of the Planck curvature anomaly. Consequently, also the lensing amplitude is now in agreement with the $\Lambda$CDM expectations.
However, the same dataset also shows other mild deviations from the baseline $\Lambda$CDM scenario~\cite{ACT:2020gnv}. In particular, the effective number of relativistic degrees of freedom in the early Universe is constrained to $N_{\rm{eff}}=2.46\pm0.26$ at $68 \%$ Confidence Level (CL hereafter); in disagreement at about two standard deviations with the reference value $N_{\rm{eff}}=3.04$ predicted by the Standard Model of particle physics for three active neutrinos~\cite{Mangano:2005cc,deSalas:2016ztq,Akita:2020szl,Froustey:2020mcq,Bennett:2020zkv,Archidiacono:2011gq}. Furthermore, also the running of the spectral index $d n_s/d\log k$, usually expected to be of order $\sim (1-n_s)^2 \lesssim 10^{-3}$ in single-field slow-roll inflation~\cite{Martin:2013tda,Akrami:2018odb}, turns out to be slightly anomalous as a mild preference for a positive running of $dn_s/d\log k=0.0128\pm0.081$ at $68 \%$ CL is observed~\cite{ACT:2020gnv,Forconi:2021que}. In addition, the constraints on the neutrino mass are quite relaxed with respect to the Planck limits, even suggesting a non vanishing $\sum m_{\nu}\sim 0.7$ eV at about one standard deviation, see also Ref~\cite{DiValentino:2021imh}. Finally, there is evidence for Early Dark Energy at more than 99\% CL~\cite{Hill:2021yec,Poulin:2021bjr}.

The question we would like to address in this paper is whether by moving to an extended parameter-space these "new" Planck-independent anomalies present in the CMB experiments, rather than be discrepant with other non-CMB observations, remain a robust predication of the CMB data and if they have any impact on the constraints of the extended scenarios. In other words, moving to extended cosmological models where all the unknown parameters are let to vary freely, does the evidence for a flat Universe, massive neutrinos, a smaller number of relativistic degrees of freedom and a positive running of the inflationary spectral index persist? Are the predictions of the different  CMB experiments consistent?  Notice that an important point to keep in mind is that in this case the constraints are obtained after marginalization over the nuisance but also "anomalous" parameters and therefore provide more robust determination with respect to the results inferred with the extra parameters fixed at $\Lambda$CDM. In light of the above mentioned cosmological tensions, answering these questions represent a crucial step both for theoretical and observational cosmology.

Our paper is organized as follows: in \autoref{sec.method} we describe both the additional parameters introduced in the sample and our data-analysis methodology; in \autoref{sec.results} we present a systematic analysis of the results model by model and parameter by parameter; finally, in \autoref{sec.conclusion}, exploiting the outcomes of our investigation, we draw some general conclusions about the different anomalies observed in the data, highlighting several possible hints for new physics beyond $\Lambda$CDM.

\section{Method}
\label{sec.method}
Our data-analysis method follows the same procedure already used in several previous papers for the Planck data: starting from the baseline $\Lambda$CDM cosmological model\footnote{We recall that the baseline $\Lambda$CDM model has six free-parameters: the baryon $\omega_{\rm b}\doteq \Omega_{\rm b}h^2$ and cold dark matter $\omega_{\rm c}\doteq\Omega_{\rm c}h^2$ energy densities, the angular size of the horizon at the last scattering surface $\theta_{\rm{MC}}$, the optical depth $\tau$, the amplitude of primordial scalar perturbation $\log(10^{10}A_{\rm S})$ and the scalar spectral index $n_s$.}, we proceed by including different combinations of additional parameters. For each extension to $\Lambda$CDM, we perform a Monte Carlo Markov Chain (MCMC) analysis using  the publicly available package \texttt{CosmoMC}~\cite{Lewis:2002ah,Lewis:2013hha} and computing the cosmological model exploiting the latest version of the Boltzmann code \texttt{CAMB}~\cite{Lewis:1999bs,Howlett:2012mh}. We vary the cosmological parameters in a range of external and conservative priors listed in \autoref{tab.priors} and explore the posteriors of our parameter space using the MCMC sampler developed for \texttt{CosmoMC} and tailored for parameter spaces with a speed hierarchy which also implements the "fast dragging" procedure described in Ref.~\cite{Neal:2005}. The convergence of the chains obtained with this procedure is tested using the Gelman-Rubin criterion~\cite{Gelman:1992zz} and we choose as a threshold for chain convergence $R-1 \lesssim 0.02 $. As is common in the literature,  the marginalized posteriors are obtained for each parameter, confidence levels are derived by symmetrically integrating the posterior around the best fit. We like to stress that the confidence levels at one standard deviation reported in the tables are around the mean value and that doubling them will not immediately provide confidence levels at two standard deviations unless the posterior is well described by a Gaussian function. Since, given the parameter degeneracy in our extended analysis, this will often NOT be the case, we warn the reader to carefully carefully the confidence limits at one standard deviation reported in the tables. To improve the readability, we highlighted in boldface in the Tables the parameter constraints that are different from the standard value at more than 95\% CL.

\begin{table}[h]
\begin{center}
\renewcommand{\arraystretch}{1.5}
\begin{tabular}{c@{\hspace{1 cm}}@{\hspace{1 cm}} c}
\hline
\textbf{Parameter}                    & \textbf{Prior}\\
\hline\hline
$\Omega_{\rm b} h^2$         & $[0.005\,,\,0.1]$\\
$\Omega_{\rm c} h^2$       & $[0.001\,,\,0.99]$\\
$100\,\theta_{\rm {MC}}$             & $[0.5\,,\,10]$\\
$\tau$                       & $[0.01\,,\,0.8]$\\
$\log(10^{10}A_{\rm s})$         & $[1.61\,,\,3.91]$\\
$n_s$                        & $[0.8\,,\, 1.2]$\\
$\Omega_k$                  &  $[-0.3\,,\,0.3]$\\
$w$                  &  $[-3\,,\, 1]$\\
$\alpha_{\rm s}\equiv dn_{\rm s} / d\log k$       &  $[-1\,,\, 1]$\\
$M_{\nu}\equiv \sum m_{\nu}$ [eV]                  &  $[0.06 , 5]$\\
$N_{\rm eff}$                  &  $[0.05\,,\, 10]$\\
\hline\hline
\end{tabular}
\end{center}
\caption{List of the parameters used in the MCMC sampling and their external flat priors. For ACTPol+WMAP and SPT3G+WMAP we additionally include a Gaussian prior on $\tau = 0.065 \pm 0.015$.}
\label{tab.priors}
\end{table}

\begin{table}[h]
\begin{center}
\renewcommand{\arraystretch}{1.5}
\begin{tabular}{c@{\hspace{1 cm}}@{\hspace{1 cm}} c}
\hline
\textbf{Parameter}                    & \textbf{Fiducial value}\\
\hline\hline
$\Omega_b h^2$ & $    0.02236$\\
$\Omega_c h^2$ & $    0.1202$\\
$100\theta_{MC}$ & $ 1.04090$\\
$\tau$ & $ 0.0544$\\
$\log(10^{10}A_{\rm s})$         & $3.045$\\
$n_s$ & $ 0.9649$\\
\hline\hline
\end{tabular}
\end{center}
\caption{Fiducial values for the parameters adopted for the simulated Planck data, assuming a $\Lambda$CDM model.}
\label{fiducial}
\end{table}

\subsection{Cosmological Models}
We analyze different extended models of cosmology that differ from the baseline case for two or more additional degrees of freedom resulting from modifications to the following sectors
\begin{itemize}[leftmargin=*]
    \item[] \textit{\textbf{Curvature}}: we relax the inflationary predictions for a flat Universe~\cite{Guth:1980zm} and allow for curved background geometries parametrized by the curvature density parameter $\Omega_k\equiv 1-\Omega$ defined in such a way that negative (positive) values correspond to a spatially closed (opened) Universe. Notice also that while the vast majority of inflationary models naturally predict spatial flatness (\ie $\Omega_k=0$), inflation in a curved Universe has been largely discussed in literature (see, \textit{e.g.}, Refs~\cite{Linde:1995xm,Linde:2003hc,Ratra:2017ezv,Bonga:2016iuf,Handley:2019anl,Bonga:2016cje,Ooba:2017ukj,Ellis:2001ym,Uzan:2003nk,Unger:2018oqo,Gordon:2020gel,Sloan:2019jyl,Forconi:2021que} ) and models with $\Omega_k<0$ can be build, as well.
    
    \item[] \textit{\textbf{Dark Energy}}: we relax the assumption $w=w_{\Lambda}\equiv-1$ for Dark Energy equation of state, leaving instead $w$ a free parameter. We refer to this class of models as $w$CDM. 
    
    \item[] \textit{\textbf{Inflation}}: we relax the assumption of scale-invariant spectral index of primordial perturbations introducing a weak scale-dependence in the primordial spectrum modeled by the running of the scalar tilt $\alpha_s \equiv dn_s/d\log k$ that quantifies the rate of change of $n_{\rm s}$ per Hubble time (we recall that  $d/d\log k = 1/H\, d/dt$ during inflation~\cite{Martin:2013tda}). In this case we parametrize the primordial scalar spectrum as 
    \begin{equation}
    \log\mathcal{P}_{\rm s}(k)=\log A_{\mathrm{s}}+\left(n_{\mathrm{s}}-1\right) \log\left(k / k_{*}\right) + \frac{\alpha_s}{2} \log^2 \left(k / k_{*}\right)
    \end{equation}
    where $k_{*}$ denotes an arbitrary pivot scale that we choose to be $k_{*}=0.05\,\rm{Mpc}^{-1}$ along all this work.
    
    \item[] \textit{\textbf{Neutrinos}}: as robustly indicated by oscillation experiments~\citep{deSalas:2020pgw,DeSalas:2018rby}, we consider neutrinos as massive particles, leaving their total mass $M_{\nu}\equiv \sum m_{\nu}$ a free parameter of the model. Indeed, while current oscillation experiments have provided a conclusive evidence for a non-vanishing mass by measuring the mass difference between neutrino flavours, its total value remains unknown and laboratory experiments only place a lower bound $\sum m_{\nu}\gtrsim 0.05$ eV \cite{Capozzi:2017ipn}. Cosmology provides a powerful (although indirect) mean to constrain their mass (see, \textit{e.g.}, Refs~\cite{deSalas:2018bym,Hagstotz:2020ukm,Vagnozzi:2019utt,Vagnozzi:2018pwo,Vagnozzi:2018jhn,Vagnozzi:2017ovm,Giusarma:2016phn,Bond:1980ha,Doroshkevich:1980zs,Capozzi:2021fjo,DiValentino:2021imh,RoyChoudhury:2018gay,RoyChoudhury:2019hls, SimonsObservatory:2019qwx,Dvorkin:2019jgs,Abazajian:2022ofy}).
    
    \item[] \textit{\textbf{Dark Radiation}}: we leave the effective number of relativistic degrees of freedom $N_{\rm eff}$ a free parameter of the cosmological model. We recall that it is defined by the relation 
    \begin{equation}
    \rho_{\mathrm{rad}}=\left[1+\frac{7}{8}\left(\frac{4}{11}\right)^{4 / 3}     N_{\mathrm{eff}}\right] \rho_{\gamma}~,
    \end{equation}
    with $\rho_{\gamma}$ the present Cosmic Microwave Background (CMB) energy-density and that it counts the effective number of relativistic particles at recombination. In the case of three active massless neutrinos, the Standard Model of particle physics predicts $N_{\rm eff}=3.044$~\cite{Mangano:2005cc,deSalas:2016ztq,Akita:2020szl,Froustey:2020mcq,Bennett:2020zkv,Archidiacono:2011gq}. However, larger (smaller) values are possible if additional (less) relativistic degrees of freedom are present in the early Universe (see, \textit{e.g.}, Refs~\cite{DiValentino:2011sv,DiValentino:2013qma,DiValentino:2015wba,Giare:2020vzo,Giare:2021cqr,DEramo:2022nvb,Baumann:2016wac,Gariazzo:2015rra,Archidiacono:2022ich,An:2022sva, Abdalla:2022yfr,SimonsObservatory:2019qwx,Green:2019glg,Dvorkin:2022jyg}).
\end{itemize}

Notice that, while there is not a reason to prefer any particular extended model, studying all the possible combinations arising from these extra parameters would be too expensive in terms of computational time. Therefore we will focus only on a sample of 8 extended models with up to 11 free parameters (5 more than the baseline case) which is large enough to allow us to derive some general reliable conclusions.

\subsection{Data}
Our baseline datasets consist of the observations of the Cosmic Microwave Background provided by the following independent experiments:
\begin{itemize}[leftmargin=*]
	\item[] \textit{\textbf{Planck 2018}} temperature and polarization (TT TE EE) likelihood~\cite{Aghanim:2019ame,Aghanim:2018eyx,Akrami:2018vks}, which also includes low multipole data ($\ell < 30$). We refer to this combination as "Planck (TT TE EE)."
	\item[] \textit{\textbf{Atacama Cosmology Telescope}} DR4 likelihood~\cite{ACT:2020frw}, combined with WMAP 9-years observations data~\cite{Hinshaw:2012aka} and a Gaussian prior on $\tau = 0.065 \pm 0.015$, as done in Ref~\cite{Aiola:2020azj}. We refer to this dataset combination as "ACTPol+WMAP."
	\item[] \textit{\textbf{South Pole Telescope}} polarization measurements SPT-3G~\cite{SPT-3G:2021eoc} combined with WMAP 9-years observations data~\cite{Hinshaw:2012aka} and a Gaussian prior on $\tau = 0.065 \pm 0.015$. We refer to this dataset combination as "SPT-3G+WMAP."
\end{itemize}

In addition, we simulate Planck temperature and polarization data assuming a best-fit flat $\Lambda$CDM model, reported in \autoref{fiducial}, and experimental noise properties similar to those reported in the Planck 2018 release, using the same methodology of Refs.~\cite{DiValentino:2019qzk,DiValentino:2020leo}. 
In other words, we compute the theoretical CMB temperature and polarization angular power spectra $C_\ell$'s with \texttt{camb}~\cite{Lewis:1999bs}, and we assume an instrumental noise $N_\ell$ such as

\begin{equation}
N_\ell = w^{-1}\exp(\ell(\ell+1)\theta^2/8\ln2)~,
\end{equation}
where $\theta$ is the experimental FWHM angular resolution of the beam and $w^{-1}$ is experimental sensitivity of Planck, as presented in the Planck legacy release~\cite{Akrami:2018vks}.

For each model, we perform a MCMC analysis over these simulated data to forecast the expected bounds on parameters and their correlations. In this way, we can evaluate any possible bias due to the large volume of the parameter space (that we shortly indicate as "volume effect" hereafter) and distinguish actual anomalies from sample bias effects. We refer to the simulated data as "Planck (forecasts)".

Finally, we often test the robustness of our results by combining the above mentioned measurements of the Cosmic Microwave Background with the following CMB-independent astrophysical observations: 

\begin{itemize}[leftmargin=*]
	\item[] \textit{\textbf{Baryon acoustic oscillations}} (BAO) measurements extracted from data from the 6dFGS~\cite{Beutler:2011hx}, SDSS MGS~\cite{Ross:2014qpa} and BOSS DR12~\cite{Alam:2016hwk} surveys. We refer to this dataset combination as "BAO."
	
    \item[] \textit{\textbf{Type Ia Supernovae}} (SNeIa) distance moduli measurements from the Pantheon sample \citep{Pan-STARRS1:2017jku}. We refer to this dataset as "Pantheon".
\end{itemize}

\section{Results}
\label{sec.results}
In this section we present and discuss in a systematic way the results obtained following the methodology outlined in \autoref{sec.method}. In particular, we dedicate a different subsection to each model analyzed in the work, providing always a table with the final results on the cosmological parameters at 68\% CL and a triangular plot with the 68\% and 95\% CL marginalized probability contours. In all the tables, together with the constraints on the free parameters, we include also the results for the  Hubble constant $H_0\,(=100\,h\,\rm{Km/s/Mpc})$; the parameter $\sigma_8$ that quantifies the amplitude of the power spectrum on scales of $8\,h^{-1}\,\rm{Mpc}$; the present day matter density parameter $\Omega_m$; the combination $S_{8} \equiv \sigma_{8} \sqrt{\Omega_{m} / 0.3} $; and the comoving sound horizon at the end of the baryonic-drag-epoch $r_{\rm drag}$. Notice that all these derived quantities are directly connected to different cosmological tensions and anomalies~\cite{DiValentino:2020zio,DiValentino:2020vvd,Abdalla:2022yfr}. Hence it is particularly interesting to study how the differences among independent datasets and experiments are recast into them. For this reason, in each subsection, we provide a concise discussion of the most interesting findings, listing and interpreting the results obtained for the relevant parameters.


\subsection{\boldmath{$\Lambda$}\textbf{CDM}  \boldmath{$+ \Omega_k + \sum m_{\nu}$}}
\label{sec.LCDM+Omk+Mnu}

\begin{table*}[htbp!]
\begin{center}
\renewcommand{\arraystretch}{1.5}
\resizebox{0.7 \textwidth}{!}{\begin{tabular}{c  |c c | c  | c   }

	
	\hline
	\textbf{Parameter} & \textbf{\nq{Planck\\ (LCDM forecasts)}}  & \textbf{\nq{Planck 2018\\ (TT TE EE)}}  &\textbf{ACTPol+WMAP}  & \textbf{SPT-3G+WMAP} \\
	\hline\hline
	$\Omega_{\rm b} h^2$ &$0.02229\pm 0.00017$& $0.02253 \pm 0.00018$ & $0.02222\pm 0.00021$&$0.02250\pm 0.00025$\\
	
	$\Omega_{\rm c} h^2$ &$0.1204 \pm 0.0017$ & $0.1184 \pm 0.0015$ &$0.1220 \pm 0.0029$&$0.1172 \pm 0.0031$\\
	
	$100\,\theta_{\rm {MC}}$ &$1.04069 \pm 0.00041$& $1.04097 \pm 0.00035$ & $1.04130 \pm 0.00066$&$1.03930 \pm 0.00071$ \\
	
	$\tau$   &$0.052 \pm 0.011$& $0.0480 \pm 0.0082$  & $0.062 \pm 0.013$ & $0.063 \pm 0.013$\\
	
	$\log(10^{10}A_{\rm S})$ &$3.040 \pm 0.021$& $3.027 \pm 0.017$ & $3.069 \pm 0.026$&$3.050 \pm 0.026$\\
	
	$n_s$ &$0.9633 \pm 0.0047$& $0.9688\pm0.0050$& $0.9658 \pm 0.0074$&$0.9677 \pm 0.0084$ \\
	
	$\Omega_k$ &$-0.020^{+0.026}_{-0.013}$&$\mathbf{-0.077^{+0.041}_{-0.021}[^{+0.058}_{-0.070}]}$ & $-0.029^{+0.035}_{-0.015}$&$-0.0009^{+0.017}_{-0.0095}$\\
	
	$\sum m_{\nu}$ [eV] &$ < 0.563$&$<0.494$ &$\mathbf{0.82^{+0.34}_{-0.31}[^{+0.58}_{-0.67}]}$ &$0.48^{+0.14}_{-0.39}$ \\
	
	$H_0$ [Km/s/Mpc] &$58  \pm 7$&$48\pm 5$  &$54^{+7}_{-8}$&$65.2 \pm 7.2$  \\
	
	$\sigma_8$&$0.721^{+0.080}_{-0.059}$&$0.687^{+0.073}_{-0.046}$  & $0.664^{+0.052}_{-0.071}$ &$
	0.714^{+0.065}_{-0.049} $\\
	
	$S_8$ &$0.870 \pm 0.055$&$1.010 \pm 0.051$ &$0.884 \pm 0.072$ &$0.766 \pm 0.060$ \\
	
	$\Omega_m$&$0.455^{+0.075}_{-0.16}$&$0.67 ^{+0.10}_{-0.20}$ &$0.56^{+0.10}_{-0.20}$&$0.355^{+0.054}_{-0.10}$ \\
	
	$r_{\rm drag}$ [Mpc] &$146.92 \pm 0.40$&$147.23 \pm 0.33$  &$146.33 \pm 0.76$&$147.53 \pm 0.77$\\
	\hline \hline
	
	\end{tabular}}
\end{center}
\caption{Constraints on cosmological parameters at 68\% CL [95\% CL] for $\Lambda$CDM$+ \Omega_k + \sum m_{\nu}$.}
\label{tab.LCDM+Omk+Mnu}
\end{table*}

\begin{figure}
\centering
\includegraphics[width=0.7\textwidth]{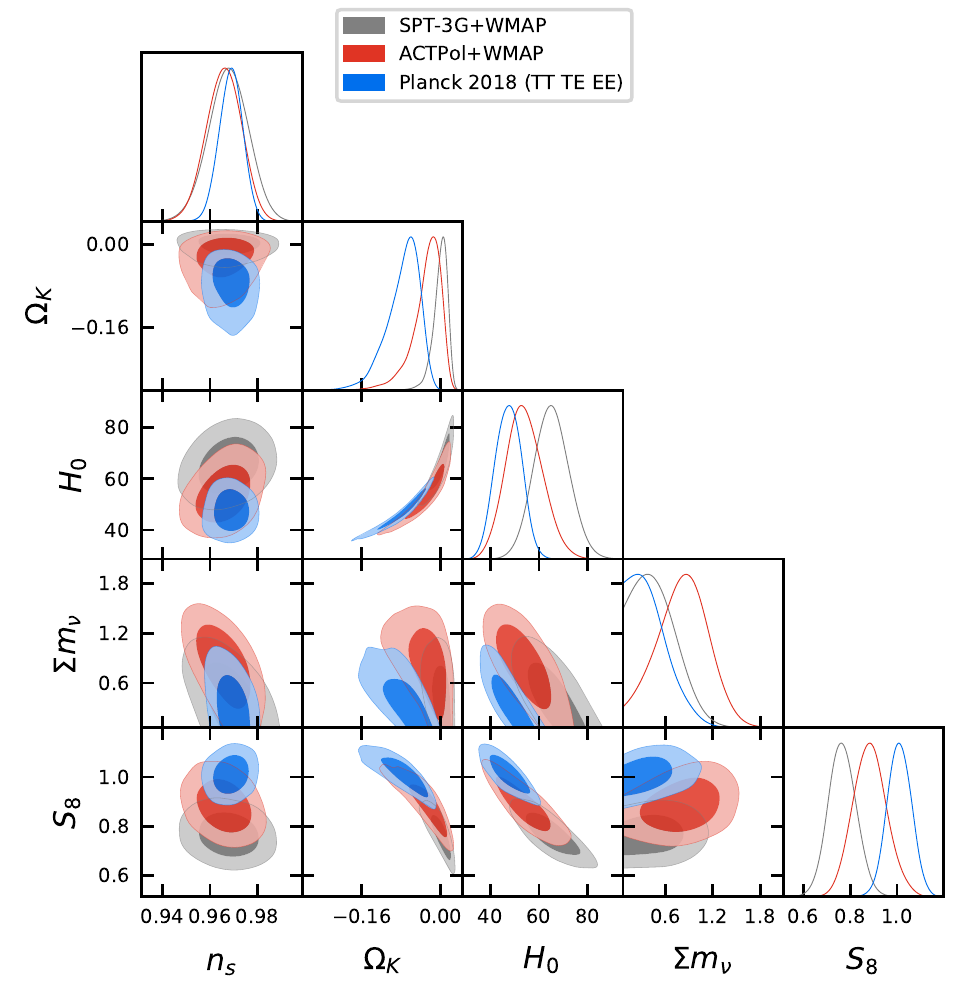}
\caption{Triangular plot showing the 1D posterior distributions and 2D contour plots for some of the parameters in $\Lambda$CDM$+ \Omega_k + \sum m_{\nu}$.}
\label{fig:omegakmnu}
\end{figure}

We start by analyzing an extension to $\Lambda$CDM which involves the spatial curvature parameter $\Omega_k$ and the total neutrino mass $\sum m_{\nu}$ as additional degrees of freedom. The numerical constraints on the parameters are provided in~\autoref{tab.LCDM+Omk+Mnu}, while the correlation plots in~\autoref{fig:omegakmnu}. Please note the significant non-gaussianity of several posteriors in~\autoref{fig:omegakmnu} that suggest a careful interpretation of the errors at one standard deviation in~\autoref{tab.LCDM+Omk+Mnu}.

A detailed analysis of the results obtained for the different experiments leads us to draw the following conclusions.

\begin{itemize}
	
\item [$H_0$)] Both Planck (TT TE EE) and ACTPol+WMAP prefer low expansion rate with the final constraints reading $H_{0}=48\pm 5$ Km/s/Mpc and $H_{0}=54^{+7}_{-8}$ Km/s/Mpc, respectively. These values are smaller compared to the results usually obtained within the baseline case and so, despite the large errors, they are both in tension with the direct measurement of the Hubble constant recently provided by the SH0ES collaboration ($H_0=73\pm1$ Km/s/Mpc); respectively at 4.9$\sigma$ and 2.7$\sigma$. However, they are both in line with our forecasts for Planck-like experiments ($H_{0}=58\pm 7$ Km/s/Mpc), providing evidence that the values of $H_0$ inferred by CMB data are largely sensitive to the underlying cosmological model and its assumptions. As concerns SPT-3G+WMAP , this is the only dataset that still predicts $H_{0}=65.2\pm 7.2$ Km/s/Mpc, close to the usual $\Lambda$CDM result. Due to the large error-bars, this value is compatible with SH0ES, as well.

\item[$\Omega_k$)] The Planck (TT TE EE) data show a definite preference for a closed Universe ($\Omega_k<0$) at more than 99\% CL, \textit{i.e.}, $-0.20<\Omega_k<-0.01$ (please note the significant non gaussianity of the posterior). However, this preference is not observed in the forecasted results for the simulated data where flatness is in fact recovered well within 1$\sigma$. Hence, we conclude that it does not reflect any sample volume effect, but it is related to genuine features in the Planck data, which remain essentially as unexplained within the baseline $\Lambda$CDM model of cosmology. Conversely, both ACTPol+WMAP ($\Omega_k=-0.029^{+0.035}_{-0.015}$) and SPT-3G+WMAP ($\Omega_k=-0.0009^{+0.017}_{-0.0095}$) are perfectly consistent with the inflationary prediction for a flat Universe ($\Omega_k=0$).

\item[$S_8$]) We observe some tensions also for the values of the parameter $S_8$ inferred by the different experiments. In particular, the result derived for Planck (TT TE EE)  ($S_8=1.010\pm 0.051$) shows a mild 1.5$\sigma$ discrepancy with the value obtained analyzing ACTPol+WMAP ($S_8=0.884\pm 0.072$) while it is in 3$\sigma$ tension with SPT-3G+WMAP ($S_8=0.766\pm 0.060$). Notice also that both ACTPol+WMAP and SPT-3G+WMAP are consistent with values $S_8 \sim 0.78$ suggested by cosmic shear surveys (such as KiDS-1000~\cite{Heymans:2020gsg,KiDS:2020ghu} or DES-y3~\cite{DES:2021vln,DES:2022ygi}). Planck real data strongly prefer a larger $S_8$, in disagreement with cosmic shear surveys measurements and SPT-3G+WMAP.

\item[$M_{\nu}$)] Analyzing the Planck (TT TE EE) data we find no evidence for a neutrino mass and the final upper bound reads $\sum m_{\nu}<0.494$ eV. This limit is slightly more constraining than our forecasted one ($\sum m_{\nu}<0.563$ eV). This is due to the excess of lensing measured by Planck, together with a combined effect of the Planck data preferences for a small $H_0$ and $\Omega_k<0$. On the other hand, ACTPol+WMAP shows a 2.6$\sigma$ preference for a non-vanishing total neutrino mass ($\sum m_{\nu}=0.82^{+0.34}_{-0.31}$ eV), while for SPT-3G+WMAP such evidence is reduced to less than 1.5$\sigma$ ($\sum m_{\nu}=0.48^{+0.14}_{-0.39}$ eV).\footnote{See also~\cite{DiValentino:2021imh}.} In any case, neither of them excludes values $\sum m_{\nu}\gtrsim 0.5 eV$ that are instead disfavored by Planck.

\end{itemize}


\subsection{\boldmath{$w$}\textbf{CDM}  \boldmath{$+ \Omega_k$}}
\label{sec.wCDM+Omk}

\begin{table*}[htbp!]
\begin{center}
\renewcommand{\arraystretch}{1.5}
\resizebox{0.6\textwidth}{!}{\begin{tabular}{c  |cc| c   }
	\hline
	\textbf{Parameter} & \textbf{\nq{Planck\\ (LCDM forecasts)}}  & \textbf{\nq{Planck 2018\\ (TT TE EE)}}  &\textbf{ACTPol+WMAP}   \\
	\hline\hline
	
	$\Omega_{\rm b} h^2$ &$0.02236 \pm 0.00016$&$0.02261 \pm 0.00017$ &  $0.02235 \pm 0.00022$   \\
	
	$\Omega_{\rm c} h^2$ &$0.1201 \pm 0.0017$&$0.1181 \pm 0.0015$ &  $0.1202 \pm 0.0030$  \\
	
	$100\,\theta_{\rm {MC}}$ &$1.04089 \pm 0.00037$&$1.04117 \pm 0.00032$& $1.04169 \pm 0.00070$ \\
	
	$\tau$   &$0.051 \pm 0.011$&$0.0483^{+0.0084}_{-0.0069}$ & $0.061 \pm 0.013$ \\
	
	$\log(10^{10}A_{\rm S})$ &$3.038 \pm 0.022$&$3.028^{+0.018}_{-0.015}$&  $3.064 \pm 0.026$ \\
	
	$n_s$ &$0.9650 \pm 0.0043$&$0.9708 \pm 0.0047$ & $0.9735 \pm 0.0065$ \\
	
	$w$ &$-0.82^{+0.54}_{-0.26}$&$-1.30^{+0.94}_{-0.47}$ &  $-0.70^{+0.61}_{-0.25}$ \\
	
	$\Omega_k$ &$-0.0193^{+0.030}_{-0.0097}$&$\mathbf{-0.046^{+0.039}_{-0.012}[^{+0.44}_{-0.69}]}$&  $-0.032^{+0.047}_{-0.011}$ \\
	
	$H_0$ [Km/s/Mpc] &$61^{+8}_{-20}$&$61^{+9}_{-20}$ &  $57^{+7}_{-20}$ \\
	
	$\sigma_8$&$0.758^{+0.069}_{-0.15}$&$0.835^{+0.095}_{-0.19}$ & $0.734^{+0.068}_{-0.17}$  \\
	
	$S_8$ &$0.882^{+0.092}_{-0.080}$&$0.961^{+0.097}_{-0.083}$& $0.907^{+0.11}_{-0.095}$  \\
	
	$\Omega_m$&$0.46^{+0.14}_{-0.27}$&$0.46^{+0.12}_{-0.30}$ & $0.53^{+0.16}_{-0.36}$ \\
	
	$r_{\rm drag}$ [Mpc] &$147.07 \pm 0.37$&$147.34 \pm 0.31$ & $146.98^{+0.76}_{-0.68}$\\
	\hline \hline
	
	\end{tabular}}
\end{center}
\caption{Constraints on cosmological parameters at 68\% CL [95\% CL] for $w$CDM$+ \Omega_k$.}
\label{tab.wCDM+Omk}
\end{table*}

\begin{figure}
\centering
\includegraphics[width=0.7\textwidth]{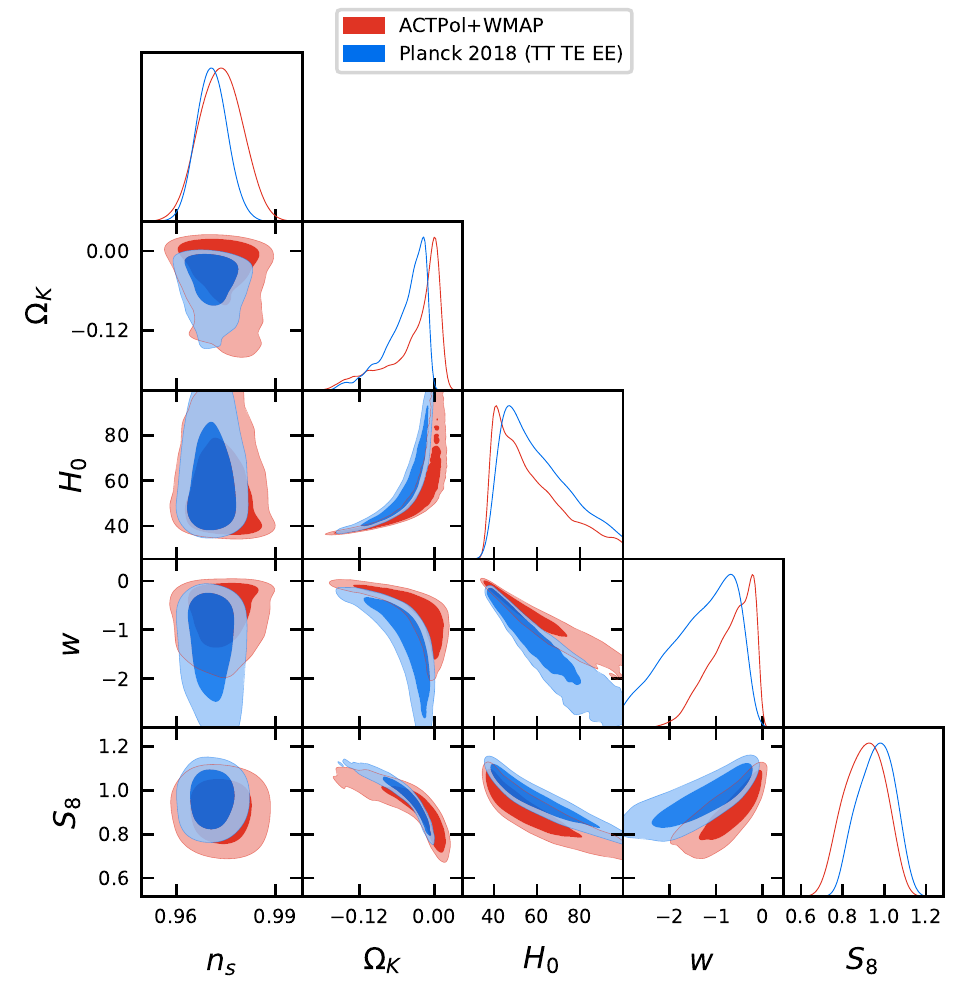}
\caption{Triangular plot showing the 1D posterior distributions and 2D contour plots for some of the parameters in $w$CDM$+ \Omega_k$.}
\label{fig:womegak}
\end{figure}

From now on we relax the assumption $w=-1$ for the Dark Energy equation of state and leave $w$ a free parameter instead. We refer to these scenarios as $w$CDM. Here, we start by analyzing the extension $w$CDM+$\Omega_k$, thus simultaneously varying $w$ and the curvature energy-density, for a total number of 8 independent parameters. The results are given in \autoref{tab.wCDM+Omk}, while the correlation plots in \autoref{fig:womegak}. 

First and foremost, we observe that when $\Omega_k$ is varied in extended parameter spaces $w$CDM, the dataset SPT-3G+WMAP is no more able to produce constraints on the cosmological parameters because the combination we have chosen (to be Planck independent) has not enough constraining power, and the MCMC analysis does not converge (\textit{i.e.}, $R-1\gg0.1$). This remains true for all the $w$CDM extensions investigated in this work, regardless of the degrees of freedom introduced in the model in addition to $\Omega_k$. We will therefore limit the analysis to the Planck (TT TE EE) and ACTPol+WMAP data.

\begin{itemize}
	
\item [$H_0$)] Due to the geometrical degeneracy between the expansion rate, the Dark Energy equation of state and the curvature parameter, both Planck (TT TE EE) and ACTPol+WMAP data weakly constrain the Hubble constant to $H_0=61^{+9}_{-20}$ Km/s/Mpc and $H_0=57^{+7}_{-20}$ Km/s/Mpc, respectively. These results further reinforce our statement that high and low redshift measurements of $H_0$ can be put into agreement in extended parameter space. Notice also that the constraints obtained by the real Planck data are in perfect agreement with the forecasted ones in case of a LCDM scenario.

\item[$\Omega_k$)] While flatness is in good agreement both with ACTPol+WMAP (at $0.7\sigma$), and this is in line with our forecasts, Planck (TT TE EE) prefers a closed universe at 99\% CL (again, note that the posterior is strongly non gaussian), actually giving $-0.148<\Omega_k<0$ at 99\% CL.

\item[$w$)] Due to the large error-bars, Planck (TT TE EE) and ACTPol+WMAP are both consistent with a cosmological constant $\Lambda$ (\textit{i.e.}, $w=-1$) at $0.3\sigma$ and $1.2\sigma$, respectively. However, for the same reason, neither a quintessential equation of state ($w>-1$) nor a phantom behaviour ($w<-1$) can be ruled out. 

\item[$S_8$)] We confirm the Planck (TT TE EE) indication for larger $S_8=0.961^{+0.097}_{-0.083}$ that strongly disfavours values $S_8\sim 0.78$ suggested by cosmic shear observations assuming a $\Lambda$CDM model. This result reflects the real data preference for larger $\sigma_8=0.835^{+0.095}_{-0.19}$ and a shift towards higher value of the matter density parameter that however turns out to be poorly constrained ($\Omega_m=0.46^{+0.12}_{-0.30}$). ACTPol+WMAP is instead in agreement with Planck, giving $S_8=0.907^{+0.11}_{-0.095}$. In this case, the result is basically driven by the shift in $\Omega_m$ rather than by $\sigma_8$.

\end{itemize}


\subsection{\boldmath{$w$}\textbf{CDM}  \boldmath{$+ \Omega_k + \sum m_{\nu}$}}
\label{sec.wCDM+Omk+Mnu}

\begin{table*}[htbp!]
\begin{center}
\renewcommand{\arraystretch}{1.5}
\resizebox{0.6\textwidth}{!}{\begin{tabular}{c |c c  | c  }
	\hline
	\textbf{Parameter} & \textbf{\nq{Planck\\ (LCDM forecasts)}}  & \textbf{\nq{Planck 2018\\ (TT TE EE)}}&\textbf{ACTPol+WMAP}  \\
	\hline\hline
	$\Omega_{\rm b} h^2$ &$0.02229 \pm 0.00017$&$0.02253^{+0.00019}_{-0.00017}$  &$0.02219\pm 0.00021$  \\
	
	$\Omega_{\rm c} h^2$ &$0.1204\pm 0.0017$&$0.1182 \pm 0.0015$  & $0.1222\pm 0.0030$ \\
	
	$100\,\theta_{\rm {MC}}$ &$1.04067 \pm 0.00040$&$1.04099 \pm 0.00035$ & $1.04127\pm0.00065$\\
	
	$\tau$   &$0.052 \pm 0.010$&$0.0472 \pm 0.0083$ &$0.062 \pm 0.013$ \\
	
	$\log(10^{10}A_{\rm S})$ &$3.038 ^{+0.021}_{-0.019}$&$3.024 \pm 0.018$ & $3.069 \pm 0.026$ \\
	
	$n_s$ &$0.9631\pm 0.0046$&$0.9691 \pm 0.0051$ & $0.9655 \pm 0.0073$\\
	
	$w$ &$-1.18^{+0.85}_{-0.37}$&$-1.59^{+0.95}_{-0.75}$  &$-1.12^{+1.1}_{-0.39}$ \\
	
	$\Omega_k$ &$-0.029^{+0.032}_{-0.012}$&$\mathbf{-0.074^{+0.055}_{-0.024}[^{+0.069}_{-0.090}]}$ & $-0.041^{+0.047}_{-0.019}$\\
	
	$\sum m_{\nu}$ [eV] &$ 0.49^{+0.18}_{-0.44}$&$0.45^{+0.12}_{-0.37}$  & $\mathbf{0.85^{+0.34}_{-0.30}[^{+0.60}_{-0.64}]}$ \\
	
	$H_0$ [Km/s/Mpc] &$60^{+7}_{-20}$&$53^{+7}_{-10}$  & $53.9^{+4.7}_{-13}$\\
	
	$\sigma_8$&$0.729^{+0.075}_{-0.15}$&$0.736^{+0.078}_{-0.15}$ & $0.662^{+0.058}_{-0.13}$\\
	
	$S_8$ &$0.874^{+0.088}_{-0.071}$&$0.990^{+0.086}_{-0.068}$ & $0.895^{+0.095}_{-0.078}$\\
	
	$\Omega_m$ & $0.49^{+0.14}_{-0.29}$&$0.61^{+0.20}_{-0.31}$ & $0.62 \pm 0.26$\\
	
	$r_{\rm drag}$ [Mpc] &$146.89\pm0.41$&$147.23 \pm 0.33$ & $146.23^{+0.83}_{-0.75}$\\
	\hline \hline
	
	\end{tabular}}
\end{center}
\caption{Constraints on cosmological parameters at 68\% CL [95\% CL] for $w$CDM$+ \Omega_k + \sum m_{\nu}$.}
\label{tab.wCDM+Omk+Mnu}
\end{table*}

\begin{figure}
\centering
\includegraphics[width=0.7\textwidth]{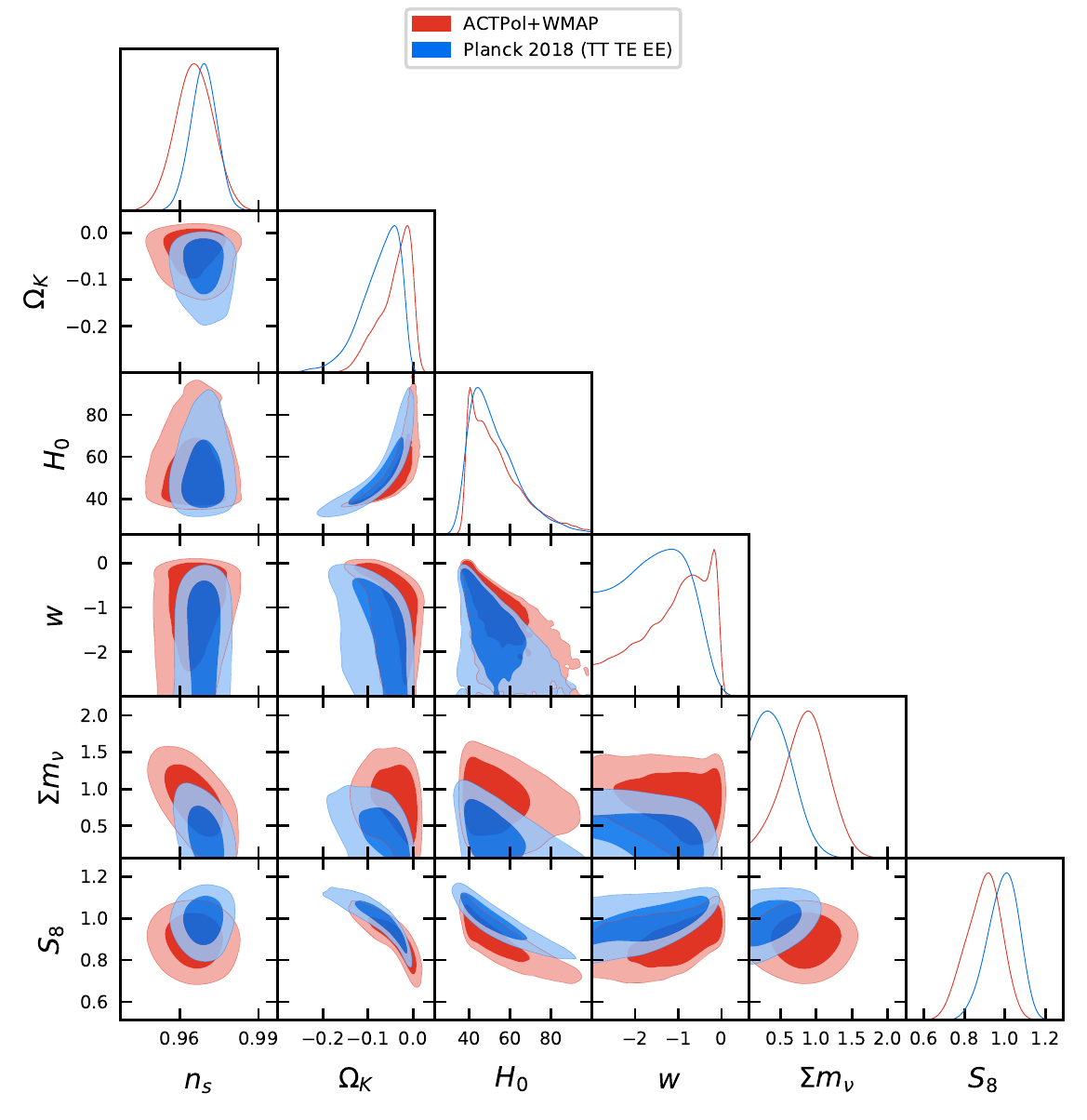}
\caption{Triangular plot showing the 1D posterior distributions and 2D contour plots for some of the parameters in $w$CDM$+ \Omega_k + \sum m_{\nu}$.}
\label{fig:womegakmnu}
\end{figure}

Starting from the previous model, we proceed by adding the total neutrino mass as an additional degree of freedom. In this case, we have 9 independent parameters. We provide the results in \autoref{tab.wCDM+Omk+Mnu}, while the correlation plots in \autoref{fig:womegakmnu}. 

As usual we summarize the most interesting findings below.

\begin{itemize}
    \item [$H_0)$] When the total neutrino mass is varied, the Planck (TT TE EE) data prefer significantly lower values of the expansion rate $H_0=53^{+7}_{-10}$ Km/s/Mpc which however remains poorly constrained. ACTPol+WMAP is in line with Planck giving $H_0=53.9^{+4.7}_{-13}$ Km/s/Mpc. Notice that, because of the large uncertainties, these results are both in agreement with SH0ES within $2\sigma$. This shift is clearly due to a volume effect because it is present also in the forecasted constraints obtained under LCDM.
    
    \item[$\Omega_k)$] Planck (TT TE EE) prefers a closed universe at more than $2\sigma$ ($\Omega_k=-0.074^{+0.069}_{-0.090}$ at 95\% CL) while ACTPol+WMAP is well within one standard deviation. Even in this extended model the forecasts show that the CMB data can still recover a flat universe within one standard deviation.
    
    \item[$w$)] Like for the previous model, all the datasets are in agreement with $w=-1$ within one standard deviation. However, given the large uncertainties, both quintessential and phantom dark energy are allowed by data.
    
    \item[$S_8$)] We confirm the Planck (TT TE EE) preference for values of $S_8 > 0.9$, with the bounds reading $S_8=0.990^{+0.086}_{-0.068}$. However, it is worth noting that when neutrinos are varied $\sigma_8=0.736^{+0.078}_{-0.15}$ is in perfect agreement with our forecasts and the $S_8$ tension becomes mostly driven by the matter density parameter that turns out to be poorly constrained and above all shifted towards higher values ($\Omega_m=0.61^{+0.20}_{-0.31}$), because of the preference for a closed universe. The same behaviour is observed for ACTPol+WMAP where we get $S_8=0.895^{+0.095}_{-0.078}$ by a combination of $\sigma_8=0.662^{+0.058}_{-0.13}$ and $\Omega_m=0.61\pm0.26$.
    
    \item[$M_{\nu})$] Analyzing the Planck (TT TE EE) data we do not find convincing evidences for a total neutrino mass since the final limit $\sum m_{\nu}=0.45^{+0.12}_{-0.37}$ eV is consistent with the forecasted result, showing that the shift is entirely due to the volume effect. On the other hand, ACTPol+WMAP show an interesting $2.8\sigma$ evidence for a non-vanishing mass $\sum m_{\nu} =0.85^{+0.34}_{-0.30}$ eV.
    
\end{itemize}

\subsection{\boldmath{$w$}\textbf{CDM}  \boldmath{$+ \Omega_k + \sum m_{\nu} + N_{\rm eff}$}}
\label{sec.wCDM+Omk+Mnu+Neff}

\begin{table*}[htbp!]
\begin{center}
\renewcommand{\arraystretch}{1.5}
\resizebox{0.6 \textwidth}{!}{\begin{tabular}{c| c c | c }
	\hline
	\textbf{Parameter} & \textbf{\nq{Planck\\ (LCDM forecasts)}}  & \textbf{\nq{Planck 2018\\ (TT TE EE)}}  &\textbf{ACTPol+WMAP}  \\
	\hline\hline
	$\Omega_{\rm b} h^2$ &$0.02224\pm 0.00032$&$0.02251 \pm 0.00026$ & $0.02152 \pm 0.00032$ \\
	
	$\Omega_{\rm c} h^2$ &$0.1200 \pm 0.0030$&$0.1183 \pm 0.0031$ &  $0.1109 \pm 0.0049$ \\
	
	$100\,\theta_{\rm {MC}}$ &$1.04072 \pm 0.00044$&$1.04098 \pm 0.00047$ &  $1.04257 \pm 0.00083$ \\
	
	$\tau$   &$0.052 \pm 0.010$&$0.0471^{+0.0082}_{-0.0073}$ & $0.056 \pm 0.013$ \\
	
	$\log(10^{10}A_{\rm S})$ &$3.037 \pm 0.023$&$3.024 \pm 0.019$  & $3.019 \pm 0.031$  \\
	
	$n_s$ &$0.961 \pm 0.012$&$0.9686 \pm 0.0095$  & $0.933^{+0.013}_{-0.015}$ \\
	
	$\Omega_k$ &$-0.031^{+0.037}_{-0.012}$&$\mathbf{-0.074^{+0.056}_{-0.024}[^{+0.069}_{-0.090}]}$  &  $-0.034^{+0.041}_{-0.015}$\\
	
	$w$ &$-1.16^{+0.90}_{-0.43}$&$-1.57^{+0.98}_{-0.77}$ &  $-1.24^{+1.1}_{-0.47}$ \\
	
	$\sum m_{\nu}$[eV] &$0.49^{+0.17}_{-0.44}$&$0.45^{+0.11}_{-0.37}$ & $\mathbf{0.84 \pm 0.25[^{+0.49}_{-0.52}]}$\\
	
	$N_{\rm eff}$ &$3.01 \pm  0.21$&$3.04 \pm 0.20$ & $\mathbf{2.32^{+0.24}_{-0.27}[^{+0.53}_{-0.48}]}$\\
	
	$H_0$ [Km/s/Mpc] &$59^{+8}_{-20}$&$53^{+6}_{-20}$ &  $52.6^{+5.0}_{-16}$\\
	
	$\sigma_8$&$0.724^{+0.078}_{-0.16}$&$0.738^{+0.076}_{-0.16}$  &  $0.641^{+0.055}_{-0.13}$ \\
	
	$S_8$ &$0.876 \pm 0.086$&$0.988^{+0.091}_{-0.070}$  & $0.856^{+0.10}_{-0.079}$ \\
	
	$\Omega_m$&$0.50^{+0.14}_{-0.31}$&$0.61^{+0.20}_{-0.33}$ & $0.60^{+0.21}_{-0.32}$ \\
	
	$r_{\rm drag}$ [Mpc] &$147.3 \pm 2.0$&$147.3 \pm 2.0$ & $154.0 \pm 3.0$\\
	\hline \hline
	
	\end{tabular}}
\end{center}
\caption{Constraints on cosmological parameters at 68\% CL [95\% CL] for $w$CDM$+ \Omega_k + \sum m_{\nu} + N_{\rm eff}$.}
\label{tab.wCDM+Omk+Mnu+Neff}
\end{table*}

\begin{figure}
\centering
\includegraphics[width=0.8\textwidth]{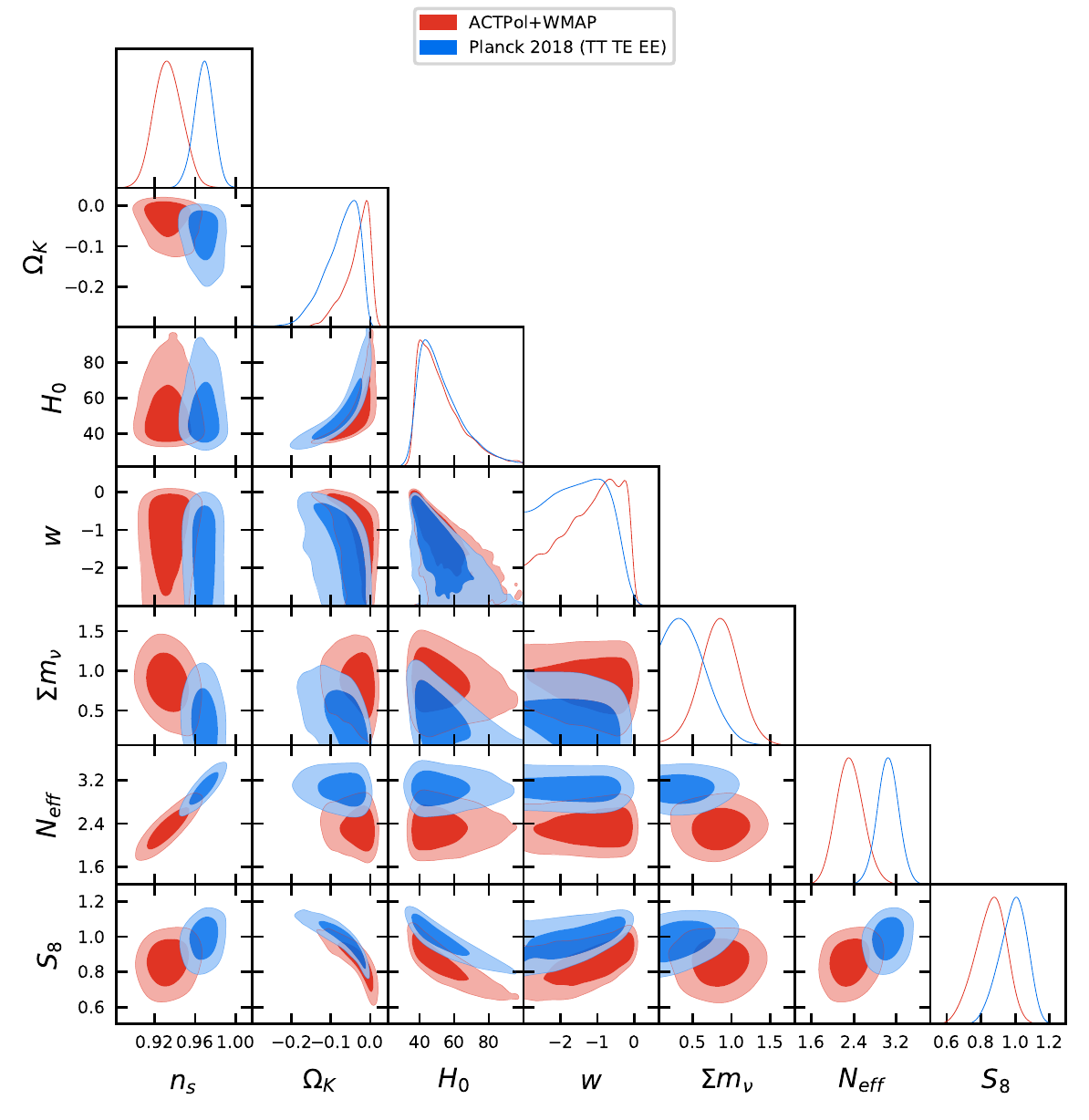}
\caption{Triangular plot showing the 1D posterior distributions and 2D contour plots for some of the parameters in $w$CDM$+ \Omega_k + \sum m_{\nu} + N_{\rm eff}$.}
\label{fig:womegakmnunnu}
\end{figure}

We further extend the previous model by adding the effective number of relativistic degrees of freedom $N_{\rm eff}$. Therefore, now we work in a 10-dimensional parameter space. In this case we summarize the results in \autoref{tab.wCDM+Omk+Mnu+Neff}, and we show the correlation plots in \autoref{fig:womegakmnunnu}. 

\begin{itemize}
    \item[$H_0)$] The inclusion of the effective number of relativistic species does not change significantly the bounds on the Hubble constant, with the final results being $H_0=53^{+6}_{-20}$ Km/s/Mpc for Planck (TT TE EE) and $H_0=52.6^{+5.0}_{-16}$ Km/s/Mpc for ACTPol+WMAP. Such values are always in agreement with SH0ES within $2\sigma$, and perfectly in agreement with the forecast results, indicating a volume effect origin. Actually, in this extended parameter space disappears the strong correlation present between $N_{\rm eff}$ and $H_0$ (see \autoref{fig:womegakmnunnu}).
    
    \item[$\Omega_k)$] The bounds on the curvature parameter are unaffected by the additional degree of freedom and the results are basically unchanged with respect to the previous case. This is because $\Omega_k$ and $N_{\rm eff}$ are not correlated, as we can see in \autoref{fig:womegakmnunnu}.
    
    \item[$w)$] Also in this case the results are basically the same obtained without $N_{\rm eff}$: all the datasets are consistent with $w=-1$ even though not enough constrained to rule out different behaviours.
    
    \item[$S_8)$] Both Planck (TT TE EE) and ACTPol+WMAP exhibit a shift towards larger matter density values that produces a mild preference for $S_8\sim 0.9$. In particular, from Planck (TT TE EE) we get $S_8=0.988^{+0.091}_{-0.070}$ given by the combination of $\Omega_m=0.61^{+0.20}_{-0.33}$ and $\sigma_8=0.738^{+0.076}_{-0.16}$, while for ACTPol+WMAP we obtain $S_8=0.856^{+0.10}_{-0.079}$ resulting from $\Omega_m=0.60^{+0.21}_{-0.32}$ and $\sigma_8=0.641^{+0.055}_{-0.13}$.
    
    \item[$M_{\nu})$] The Planck (TT TE EE) limit on the total neutrino mass remains robust under the inclusion of the effective number of relativistic neutrinos in the sample, with the final bound reading $\sum m_{\nu}=(0.45^{+0.11}_{-0.37})$ eV, clearly due to the volume effect because in agreement with the forecasted results. Conversely, ACTPol+WMAP give a slightly more tight constraint $\sum m_{\nu}=0.84\pm 0.25$ eV, further enforcing the preference for a non-vanishing mass that now reaches the level of $3.4\sigma$.
    
    \item[$N_{\rm eff})$ ] As concerns the effective number of relativistic degrees of freedom, analyzing the Planck (TT TE EE) data we find $N_{\rm eff}=3.04\pm 0.20$ in perfect agreement with the Standard Model prediction for three active massless neutrinos. On the other hand, ACTPol+WMAP gives $N_{\rm eff}=2.32^{+0.24}_{-0.27}$, showing instead a disagreement at more than 99\% CL with respect to the baseline value of particle physics and suggesting a smaller amount of radiation in the early Universe (confirming the results for the baseline case already discussed in the literature~\cite{ACT:2020gnv}). This is not a volume effect because the forecasts indicate the ability of the CMB data to recover the standard value.
    
\end{itemize}

\subsection{\boldmath{$w$}\textbf{CDM}  \boldmath{$+ \sum m_{\nu} + N_{\rm eff}  + \alpha_s$}}
\label{sec.wCDM+Mnu+Neff+nrun}

\begin{table*}[htbp!]
	\begin{center}
		\renewcommand{\arraystretch}{1.5}
		\resizebox{1 \textwidth}{!}{\begin{tabular}{c | c c c c| c  c  c  |c c c   }
				\hline
				\textbf{Parameter} & \textbf{\nq{Planck\\ (LCDM forecasts)}}  & \textbf{\nq{Planck 2018\\ (TT TE EE)}}  &  \textbf{\nq{\\ +BAO}} & \textbf{\nq{\\ +Pantheon}} & \textbf{ACTPol+WMAP}  & \nq{\\\textbf{+BAO}} & \nq{\\\textbf{+Pantheon}}  & \textbf{SPT-3G+WMAP} & \nq{\\ \textbf{+BAO}} & \nq{\\ \textbf{+ Pantheon}} \\
				\hline\hline
				
				$\Omega_{\rm b} h^2$ &$0.02219\pm0.00031$&$0.02217 \pm 0.00024$& $0.02224 \pm 0.00022$ & $0.02212 \pm 0.00023$ & $0.02149\pm 0.00036$& $0.02209 \pm 0.00034$&$0.02158\pm0.00036$   & $0.02256 \pm 0.00040$ & $0.02283 \pm 0.00035$ &$0.02258\pm 0.00039$ \\
				
				$\Omega_{\rm c} h^2$ &$0.1200 \pm 0.0037$&$0.1163 \pm 0.0033$& $0.1165 \pm 0.0033$ & $0.1162 \pm 0.0033$ & $0.1116 \pm 0.0056$& $0.1124 \pm 0.0059$&$0.1114 \pm 0.0056$   & $0.1167 \pm 0.0065$ & $0.1171 ^{+0.0062}_{-0.0072}$ &$0.1168^{+0.0061}_{-0.0070}$\\
				
				$100\,\theta_{\rm {MC}}$ &$1.04077 \pm 0.00052$&$1.04135 \pm 0.00048$& $1.04136 \pm 0.00049$ & $1.04138 \pm 0.00051$ & $1.04253^{+0.00091}_{-0.0010}$& $1.04245 \pm 0.00093$&$1.04250 \pm 0.00096$   & $1.03930 \pm 0.00090$ & $1.03939 \pm 0.00087$ &$1.03930 \pm 0.00089$\\
				
				$\tau$   &$0.0544 \pm 0.0099$&$ 0.0546 \pm 0.0080$& $0.0558 \pm 0.0082$ & $0.0548 \pm 0.0081$ & $0.059 \pm 0.012$& $0.060 \pm 0.013$&$0.059 \pm 0.013$   & $0.062 \pm 0.013$ & $0.063 \pm 0.012$ &$0.062 \pm 0.013$\\
				
				$\log(10^{10}A_{\rm S})$&$3.042 \pm 0.022$&$ 3.036\pm0.019$& $3.040\pm0.019$ & $3.037 \pm 0.019$ & $3.027 \pm 0.032$& $3.036 \pm 0.031$ & $3.027\pm0.031$   & $3.043\pm0.031$ & $3.047\pm0.030$ &$3.044 \pm 0.030$\\
				
				$n_s$ &$0.960 \pm 0.014$&$0.951 \pm 0.011$& $0.954 \pm 0.010 $ & $0.949 \pm 0.011$ & $0.934 \pm 0.022$& $0.968 \pm 0.019$&$0.938 \pm 0.022$   & $0.978 \pm 0.024$ & $0.992 \pm 0.020$ &$0.978 \pm 0.023$\\

				$w$ &$-1.48^{+0.57}_{-0.48}$&$-1.64^{+0.28}_{-0.40}$& $-1.072^{+0.079}_{-0.061} $ & $-1.064^{+0.048}_{-0.040}$ & $-1.75 \pm 0.68$& $-1.06^{+0.12}_{-0.090}$&$\mathbf{-1.25^{+0.11}_{-0.10}[^{+0.20}_{-0.21}]}$   & $-1.59^{+0.99}_{-0.57}$ & $-0.941^{+0.10}_{-0.078}$ &$-1.062^{+0.092}_{-0.061}$\\
				
				$\sum m_{\nu}$ &$< 0.449$&$< 0.139$& $< 0.0721$ & $< 0.125$ & $\mathbf{0.81 \pm 0.28[^{+0.58}_{-0.54}]}$& $< 0.271$&$\mathbf{0.71 \pm 0.28}[^{+0.53}_{-0.56}]$   & $0.48^{+0.10}_{-0.42}$ & $<0.282$ &$0.46^{+0.11}_{-0.39}$\\
				
				$N_{\rm eff}$&$2.98 \pm 0.27$&$ 2.76 \pm 0.22$& $2.81 \pm 0.22$ & $2.73 \pm 0.22$ & $2.34 \pm 0.39$& $2.75^{+0.37}_{-0.41}$ &$2.39^{+0.36}_{-0.41}$   & $3.14^{+0.43}_{-0.48}$ & $3.33^{+0.41}_{-0.49}$ &$3.14^{+0.41}_{-0.49}$\\
				
				$d n_s / d\log k$ &$0.0014 \pm 0.0093$&$ -0.0098 \pm 0.0079$& $-0.0100 \pm 0.0077$ & $-0.0117 \pm 0.0081$ & $0.002 \pm 0.013$& $0.009 \pm 0.011$&$0.002 \pm 0.013$   & $0.011 \pm 0.014$ & $0.014 \pm 0.012$ &$0.009 \pm 0.013$\\
				
				$H_0$ [Km/s/Mpc] &$77\pm 10$&$ >79.1[>65.2]$& $67.9 \pm 1.7$ & $66.3\pm1.7$ & $70^{+10}_{-20}$& $66.8^{+2.0}_{-2.4}$&$61.6^{+2.1}_{-2.6}$   & $86^{+10}_{-30}$ & $67.9^{+2.1}_{-2.4}$ &$67.3^{+2.7}_{-3.1}$\\
				
				$\sigma_8$&$0.85^{+0.12}_{-0.10}$&$ 0.948^{+0.11}_{-0.059}$& $0.821 \pm 0.020$ & $0.806^{+0.024}_{-0.016}$ & $0.75^{+0.10}_{-0.12}$& $0.770 \pm 0.034$&$0.697^{+0.038}_ {-0.048}$   & $0.85 ^{+0.11}_{-0.23}$ & $0.745 \pm 0.033$ &$0.730^{+0.049}_{-0.039}$\\
				
				$S_8$ &$0.778 \pm 0.050$&$ 0.774^{+0.031}_{-0.042}$& $0.824^{+0.015}_{-0.014}$ & $0.829 \pm 0.017$ & $0.742^{+0.060}_{-0.070}$& $0.779^{+0.035}_{-0.028}$&$0.775 \pm 0.039$   & $0.704\pm0.076$ & $0.756^{+0.032}_{-0.029}$ &$0.753 \pm 0.036$\\
				
				$\Omega_m$&$0.270^{+0.040}_{-0.12}$&$0.208^{+0.020}_{-0.069}$& $0.303 \pm 0.013$ & $0.318^{+0.012}_{-0.015}$ & $0.324^{+0.062}_{-0.17}$& $0.307 \pm 0.014$&$0.372^{+0.028}_{-0.025}$   & $0.252 ^{+0.073}_{-0.18}$ & $0.309 \pm 0.012$ &$0.320^{+0.023}_{-0.028}$\\
				
				$r_{\rm drag}$ [Mpc] &$147.6 \pm 2.5$&$ 149.9 \pm 2.2$& $149.5 \pm 2.2$ & $150.1 \pm 2.3$ & $153.8 \pm 4.1$& $151.1 \pm 4.0$&$153.7 \pm 4.1$   & $147.2 \pm 4.2$ & $146.0 \pm 4.2$ &$147.2\pm4.2$\\
				
				\hline \hline
				
		\end{tabular}}
	\end{center}
	\caption{Constraints on cosmological parameters at 68\% CL [95\% CL] for $w$CDM$+ \sum m_{\nu} + N_{\rm eff} + \alpha_s$.}
	\label{tab.wCDM+Mnu+Neff+nrun}
\end{table*}

\begin{figure}
\centering
\includegraphics[width=0.8\textwidth]{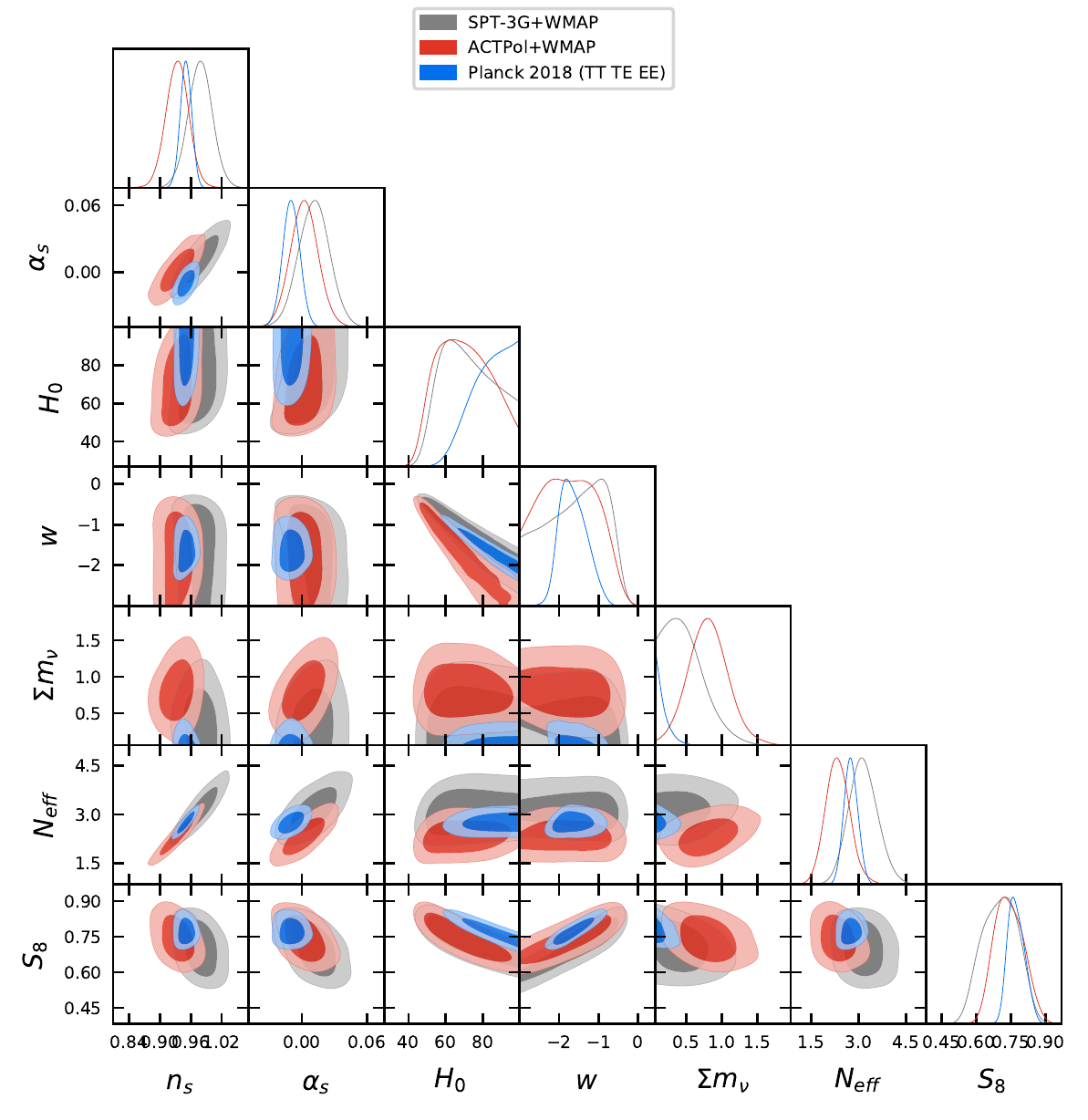}
\caption{Triangular plot showing the 1D posterior distributions and 2D contour plots for some of the parameters in $w$CDM$+ \sum m_{\nu} + N_{\rm eff} + \alpha_s$.}
\label{fig:wmnunnunrun}
\end{figure}

\begin{figure}
\centering
\includegraphics[width=0.8\textwidth]{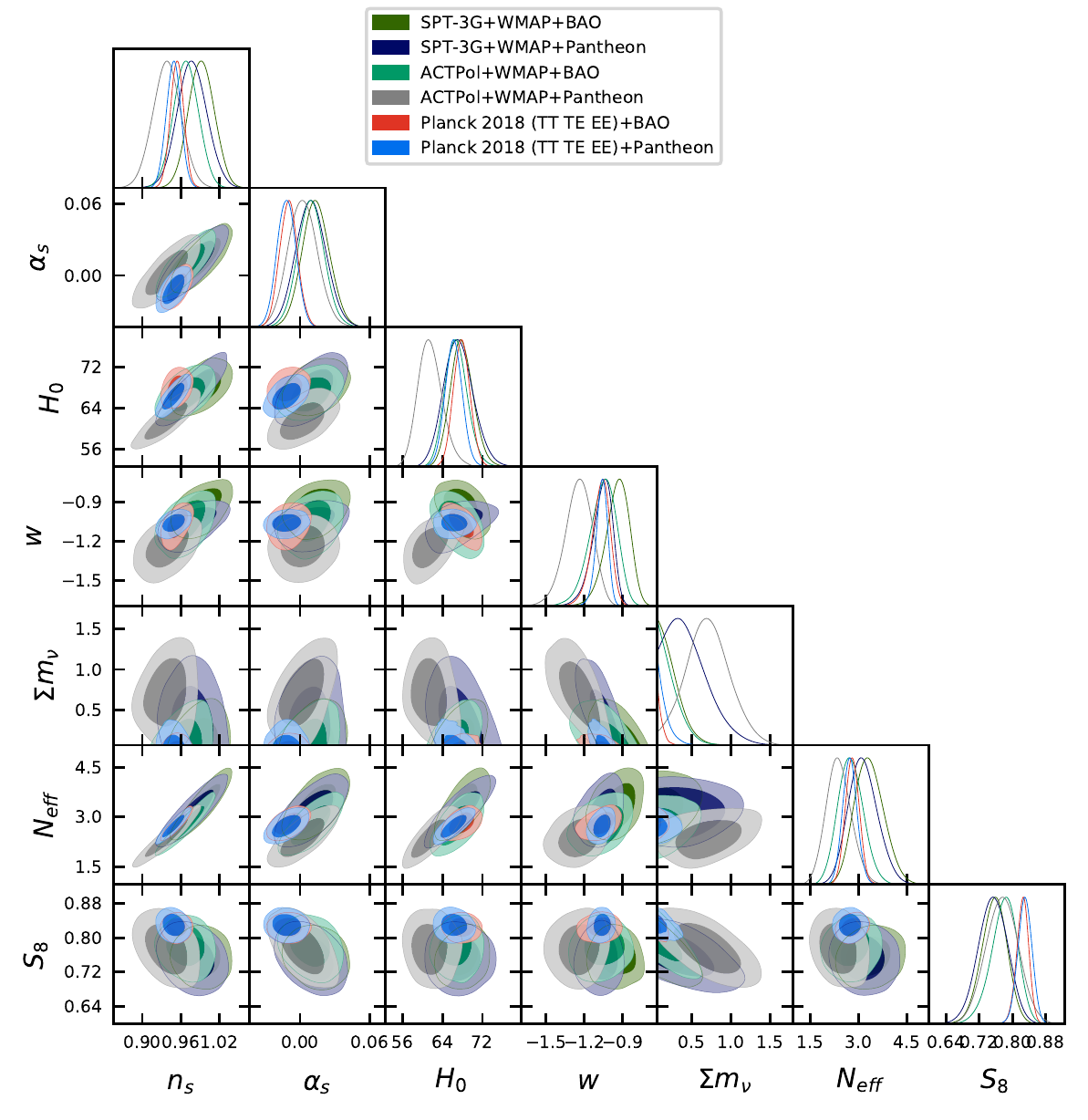}
\caption{Triangular plot showing the 1D posterior distributions and 2D contour plots for some of the parameters in $w$CDM$+ \sum m_{\nu} + N_{\rm eff} + \alpha_s$, with the inclusion of BAO and Pantheon data.}
\label{fig:wmnunnunrunBP}
\end{figure}

In this extension, we assume a flat background geometry (thus fixing $\Omega_k=0$) and replace the curvature parameter with the running of the spectral index $\alpha_s\doteq dn_s/d\log k$ remaining always in a 10-dimensional parameter-space. Furthermore, from now on, we include in the analysis also the other CMB-independent datasets listed in \autoref{sec.method} to test the robustness of the hints for new CMB spectral-anomalies discussed so far under the inclusion of astrophysical observations. We summarize the results in \autoref{tab.wCDM+Mnu+Neff+nrun}, and we show the correlation plots in \autoref{fig:wmnunnunrun} and \autoref{fig:wmnunnunrunBP}. 

\begin{itemize}
\item [$H_0$)] The Planck (TT TE EE) data are unable to place strong constraints on the Hubble constant $H_0$ because of the geometrical degeneracy with the other parameters, as shown by the forecasted data, so reducing the Hubble tension to $2\sigma$. On the other hand, when Planck is combined with BAO ($H_0=67.9\pm 1.7$ Km/s/Mpc) or Pantheon ($H_0=66.3\pm 1.7$ Km/s/Mpc) we observe a reduced tension with SH0ES at the level of 2.6$\sigma$ and $3.4\sigma$, respectively, due solely to the larger error bars. As concerns ACTPol+WMAP, we can see that the Hubble constant is still poorly constrained to $H_0=70^{+10}_{-20}$ Km/s/Mpc. Notice however that this value is in agreement with the local result. Combining ACTPol+WMAP+BAO ($H_0=66.8^{+2.0}_{-2.4}$ Km/s/Mpc) and ACTPol+WMAP+Pantheon ($H_0=61.6^{+2.1}_{-2.6}$ Km/s/Mpc) we recover the usual tensions with SH0ES (at $2.8\sigma$ and $4.9\sigma$, respectively) already observed for the Planck data. Finally, SPT-3G+WMAP give an almost unconstrained $H_0$ as the other CMB data alone in this extended scenario, and when they are combined with BAO (Pantheon) we obtain $H_0=67.9^{+2.1}_{-2.4}$ Km/s/Mpc ($H_0=67.3^{+2.7}_{-3.1}$ Km/s/Mpc), reducing the tension at the level of $2.2\sigma$ ($2\sigma$) tension, mostly due to the larger error-bars of these dataset combinations.

\item [$w$)] Planck (TT TE EE) shows a mild preference for phantom Dark Energy ($w<-1$) at the statistical level of $\sim 2\sigma$. When it is combined with BAO or Pantheon this preference disappears and $w=-1$ is recovered within $1\sigma$ and $1.3\sigma$, respectively. ACTPol+WMAP and ACTPol+WMAP+BAO are both consistent with a cosmological constant at about one standard deviation, while ACTPol+WMAP+Pantheon suggest the same $2.3\sigma$ preference for $w<-1$ already observed for Planck. Finally, all the SPT-3G+WMAP combinations are consistent with $w=-1$ at $1\sigma$.

\item [$S_8$)] In this extended framework the indication for larger values $S_8\sim 0.9 - 1$, discussed so far when the curvature is free to vary, is not observed and all the datasets are consistent with a lower $S_8\sim 0.7 - 0.8$, in agreement with the cosmic shear measurements that are considering extended scenarios beyond the standard model~\cite{KiDS:2020ghu,DES:2022ygi}. However, it is worth noting that for Planck (TT TE EE) this result recast a preference for a smaller matter density $\Omega_m=0.208^{+0.020}_{-0.069}$ and a larger $\sigma_8=0.948^{+0.11}_{-0.059}$. This is the same behaviour observed also for SPT-3G+WMAP. Instead, when BAO and Pantheon are combined with Planck, we recover the usual result $\Omega_m\sim 0.3$ and hence an indication for $\sigma_8\sim 0.8$. Finally, for ACTPol+WMAP, and its combinations, and for SPT-3G+WMAP+BAO and SPT-3G+WMAP+Pantheon we find both lower $S_8$ and $\sigma_8$ values, with the limit case of ACTPol+WMAP+Pantheon where we have $S_8=0.775\pm0.039$ and $\sigma_8=0.697^{+0.038}_{-0.048}$. These discordant behaviours are translated into tensions about the values of $\sigma_8$ inferred by the different datasets that range between 1.7$\sigma$ and 3.7$\sigma$.

 \item[$M_{\nu}$)] The Planck (TT TE EE) data constrain the total neutrino mass to $\sum m_{\nu}<0.139$ eV. This limit is a factor of 4 better than the expected result from the simulated data and this is due to the famous excess of lensing (or $A_{\rm lens}$~\cite{Calabrese:2008rt,Planck:2018vyg,Abdalla:2022yfr} anomaly) of the real Planck data~\cite{Capozzi:2017ipn,Capozzi:2021fjo,DiValentino:2021imh,DiValentino:2019dzu}. When Planck is combined with BAO (Pantheon) this limit can be substantially (slightly) improved to $\sum m_{\nu}<0.072$eV ($\sum m_{\nu}<0.125$eV), placing anyway stringent constraints on the total neutrino mass. On the other hand, ACTPol+WMAP confirm the preference for a non-vanishing total mass $\sum m_{\nu}=0.81\pm0.28$ eV, here at the level of $2.9\sigma$~\cite{DiValentino:2021imh}.  This preference is only slightly reduced to $2.5\sigma$ for ACTPol+WMAP+Pantheon ($\sum m_{\nu}=0.71 \pm 0.28$ eV) while it disappears for ACTPol+WMAP+BAO ($\sum m_{\nu}<0.271$ eV). Finally, SPT-3G+WMAP (+Pantheon) dataset combinations give evidence for a neutrino mass at slightly more than $1\sigma$, while SPT-3G+WMAP+BAO is perfectly consistent with zero.

\item[$N_{\rm eff}$)] The dataset Planck (TT TE EE) is consistent with $N_{\rm eff}=3.044$ at $1.3\sigma$ and including BAO and Pantheon does not change the result. In this case, the ACTPol+WMAP preference for less radiation discussed in the previous models is reduced to the level of $1.8\sigma$ and does not change including Pantheon. Instead, combining ACTPol+WMAP+BAO we recover the baseline value within one standard deviation. Finally, SPT-3G+WMAP, SPT-3G+WMAP+BAO and SPT-3G+WMAP+Pantheon are all in agreement with the Standard Model prediction, as well.

\item[$\alpha_s$)] All the datasets are in agreement with $d n_s / d\log k=0$ at most within $1.5\sigma$. 

\end{itemize}


\subsection{\boldmath{$w$}\textbf{CDM}  \boldmath{$+ \Omega_k + \sum m_{\nu} + \alpha_s$}}
\label{sec.wCDM+Omk+Mnu+nrun}

\begin{table*}[htbp!]
\begin{center}
\renewcommand{\arraystretch}{1.5}
\resizebox{1\textwidth}{!}{\begin{tabular}{c | c c c c | c  c  c|  c c  }
	\hline
	\textbf{Parameter} & \textbf{\nq{Planck\\ (LCDM forecasts)}}  & \textbf{\nq{Planck 2018\\ (TT TE EE)}}  & \textbf{\nq{\\ +BAO}} & \textbf{\nq{\\ +Pantheon}} &   \textbf{ACTPol+WMAP}  & \nq{\\\textbf{+BAO}} & \nq{\\\textbf{+Pantheon}} &  \textbf{\nq{\textbf{SPT-3G+WMAP} \\ + BAO}}& \textbf{\nq{\textbf{SPT-3G+WMAP} \\ +Pantheon}} \\
	\hline\hline
	
	$\Omega_{\rm b} h^2$ &$0.02222\pm 0.00021$&$0.02253\pm 0.00019$ & $0.02243 \pm 0.00016$&$0.02255\pm 0.00018$& $0.02199\pm 0.00022$ & $0.02214 \pm 0.00024$ &$0.02208\pm 0.00023$  &$0.02254\pm 0.00025
	$ &$0.02253\pm 0.00025$\\
	
	$\Omega_{\rm c} h^2$ &$0.1204 \pm 0.0017$&$0.1183 \pm 0.0016$ &$0.1198 \pm 0.0014$&$0.1186 \pm 0.0015$& $0.1182 \pm 0.0032$& $0.1195 \pm 0.0032$ &$0.1186 \pm 0.0032$& $0.1168 \pm 0.0037$ &$0.1157 \pm 0.0038$\\
	
	$100\,\theta_{\rm {MC}}$ &$1.04062 \pm 0.00040$&$1.04099 \pm 0.00035$ &$1.04095 \pm 0.00032$&$1.04107 \pm 0.00034$& $1.04152 \pm 0.00066$ & $1.04151 \pm 0.00066$ &$1.04148 \pm 0.00066$ &$1.03934 \pm 0.00071$ &$1.03939 \pm 0.00071$\\
	
	$\tau$   &$0.051 \pm 0.010$&$0.0473 \pm 0.0083$ & $0.0563 \pm 0.0081$&$0.0506 \pm 0.0082$& $0.053 \pm 0.013$ &$0.063 \pm 0.013$& $0.059 \pm 0.013$ &$0.064 \pm 0.013$&$0.061 \pm 0.013$\\
	
	$\log(10^{10}A_{\rm S})$ &$3.035 \pm 0.022$&$3.025 \pm 0.018$ & $3.049\pm0.017$&$3.034 \pm 0.017$& $3.032 \pm 0.029$ & $3.060 \pm 0.027$ &$3.048 \pm 0.028$ & $3.051 \pm 0.028$ &$3.042 \pm 0.029$\\
	
	$n_s$ &$0.9625 \pm 0.0048$&$0.9689 \pm 0.0054$ & $0.9648\pm 0.0048$&$0.9685 \pm 0.0051$& $0.9722 \pm 0.0079$ & $0.9747 \pm 0.0076$ &$0.9743 \pm 0.0078$ & $0.971 \pm 0.012$ &$0.974 \pm 0.012$\\
	
	$w$ &$-1.18^{+0.92}_{-0.43}$&$-1.55^{+1.0}_{-0.75}$ & $-1.038^{+0.098}_{-0.088}$&$\mathbf{-1.27^{+0.14}_{-0.087}[^{+0.22}_{-0.25}]}$& $-1.47^{+1.1}_{-0.70}$ & $-0.99^{+0.13}_{-0.11}$ &$-1.29^{+0.18}_{-0.12}$ & $0.892^{+0.11}_{-0.091}$ &$-1.08^{+0.14}_{-0.076}$\\
	
	$\Omega_k$ &$-0.033^{+0.038}_{-0.013}$&$\mathbf{-0.074^{+0.058}_{-0.025}[\pm0.068]}$ & $0.0003^{+0.0027}_{-0.0037}$&$\mathbf{-0.029^{+0.011}_{-0.010}[^{+0.019}_{-0.020}]}$& $-0.060^{+0.052}_{-0.024}$ & $0.0106^{+0.0058}_{-0.0069}$ &$-0.010 \pm 0.012$ & $0.0092^{+0.0056}_{
-0.0072}$ &$-0.001^{+0.014}_{-0.011}$\\
	
	$\sum m_{\nu}$[eV] &$0.57^{+0.24}_{-0.47}$&$0.43^{+0.16}_{-0.37}$ & $< 0.0799$&$< 0.209$& $\mathbf{1.17 \pm 0.31[^{+0.60}_{-0.67}]}$& $0.61^{+0.22}_{-0.48}$ &$\mathbf{0.86^{+0.34}_{-0.30}[^{+0.60}_{-0.65}]}$ & $< 0.422$ &$0.46^{+0.14}_{-0.37}$\\
	
	$\alpha_s$ &$0.0034 \pm 0.0070$&$-0.0005 \pm 0.0067$ & $-0.0054 \pm 0.0068$&$-0.0023 \pm 0.0065$& $\mathbf{0.0238 \pm 0.0088[\pm0.017]}$& $0.0143 \pm 0.0090$ &$\mathbf{0.0195 \pm 0.0089[\pm0.017]}$ & $0.002 \pm 0.011$ &$0.008 \pm 0.012$\\
	
	$H_0$ [Km/s/Mpc] &$59^{+8}_{-20}$&$53.2^{+5.8}_{-16}$ & $68.6^{+1.5}_{-1.8}$&$60.5 \pm 2.5$& $50.9^{+5.3}_{-14}$ & $67.4^{+1.7}_{-2.0}$ &$62.6^{+2.8}_{-3.6}$ &$66.1^{+1.5}_{-1.8}$ &$66.5\pm3.9$\\
	
	$\sigma_8$&$0.712^{+0.076}_{-0.16}$&$0.742^{+0.075}_{-0.16}$ & $0.821 \pm 0.027$&$0.812^{+0.031}_{-0.018}$& $0.611^{+0.053}_{-0.11}$ & $0.711 \pm 0.054$ &$0.693^{+0.042}_{-0.058}$ & $0.713^{+0.042}_{-0.035}$ &$0.725^{+0.049}_{-0.038}$\\
	
	$S_8$ &$0.875 ^{+0.089}_{-0.080}$&$0.989^{+0.095}_{-0.063}$ & $0.826 \pm 0.016$&$0.927 \pm 0.037$& $0.872^{+0.092}_{-0.076}$& $0.742^{+0.056}_{-0.048}$ &$0.784 \pm 0.052$ & $0.745^{+0.039}_{-0.030}$ &$0.754 \pm 0.046$\\
	
	$\Omega_m$&$0.52^{+0.15}_{-0.31}$&$0.61^{+0.21}_{-0.34}$ & $0.305 \pm 0.016$&$0.393^{+0.030}_{-0.036}$& $0.68^{+0.25}_{-0.30}$ & $0.327 \pm 0.018$ &$0.386 \pm 0.042$ & $0.328 \pm 0.017$ &$0.327^{+0.033}_{-0.044}$\\
	
	$r_{\rm drag}$ [Mpc] &$146.89\pm 0.41$&$147.23 \pm 0.35$ & $147.08 \pm 0.32$&$147.24 \pm 0.32$&  $147.23 \pm 0.84$ & $147.22 \pm 0.83$ &$147.33 \pm 0.84$ & $147.68 \pm 0.93$ &$147.92 \pm 0.96$\\
	\hline \hline
	
	\end{tabular}}
\end{center}
\caption{Constraints on cosmological parameters at 68\% CL [95\% CL] for $w$CDM$+ \Omega_k + \sum m_{\nu} + \alpha_s$.}
\label{tab.wCDM+Omk+Mnu+nrun}
\end{table*}

\begin{figure}
\centering
\includegraphics[width=0.8\textwidth]{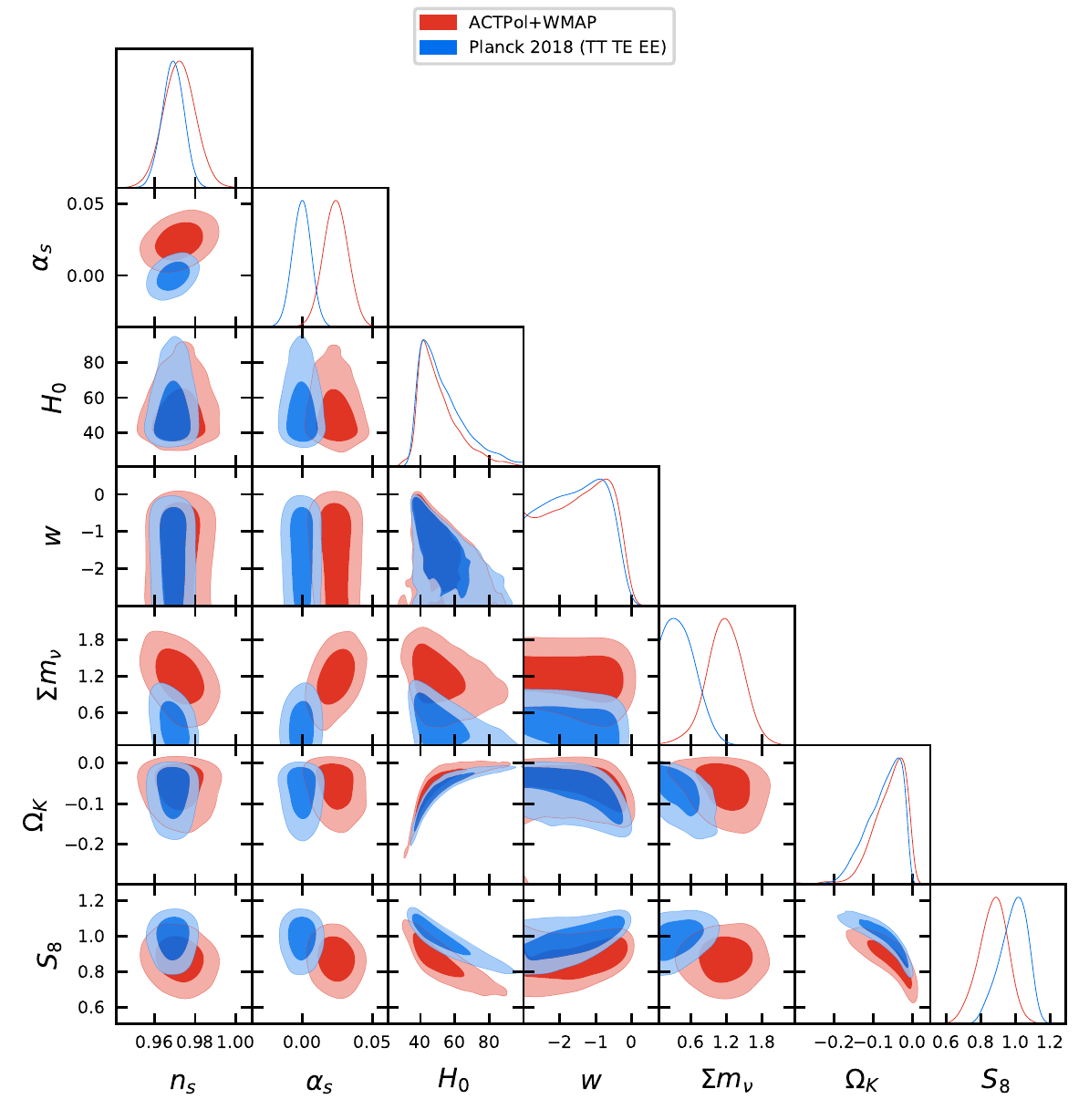}
\caption{Triangular plot showing the 1D posterior distributions and 2D contour plots for some of the parameters in $w$CDM$+ \Omega_k + \sum m_{\nu} + \alpha_s$.}
\label{fig:womegakmnunrun}
\end{figure}

\begin{figure}
\centering
\includegraphics[width=0.8\textwidth]{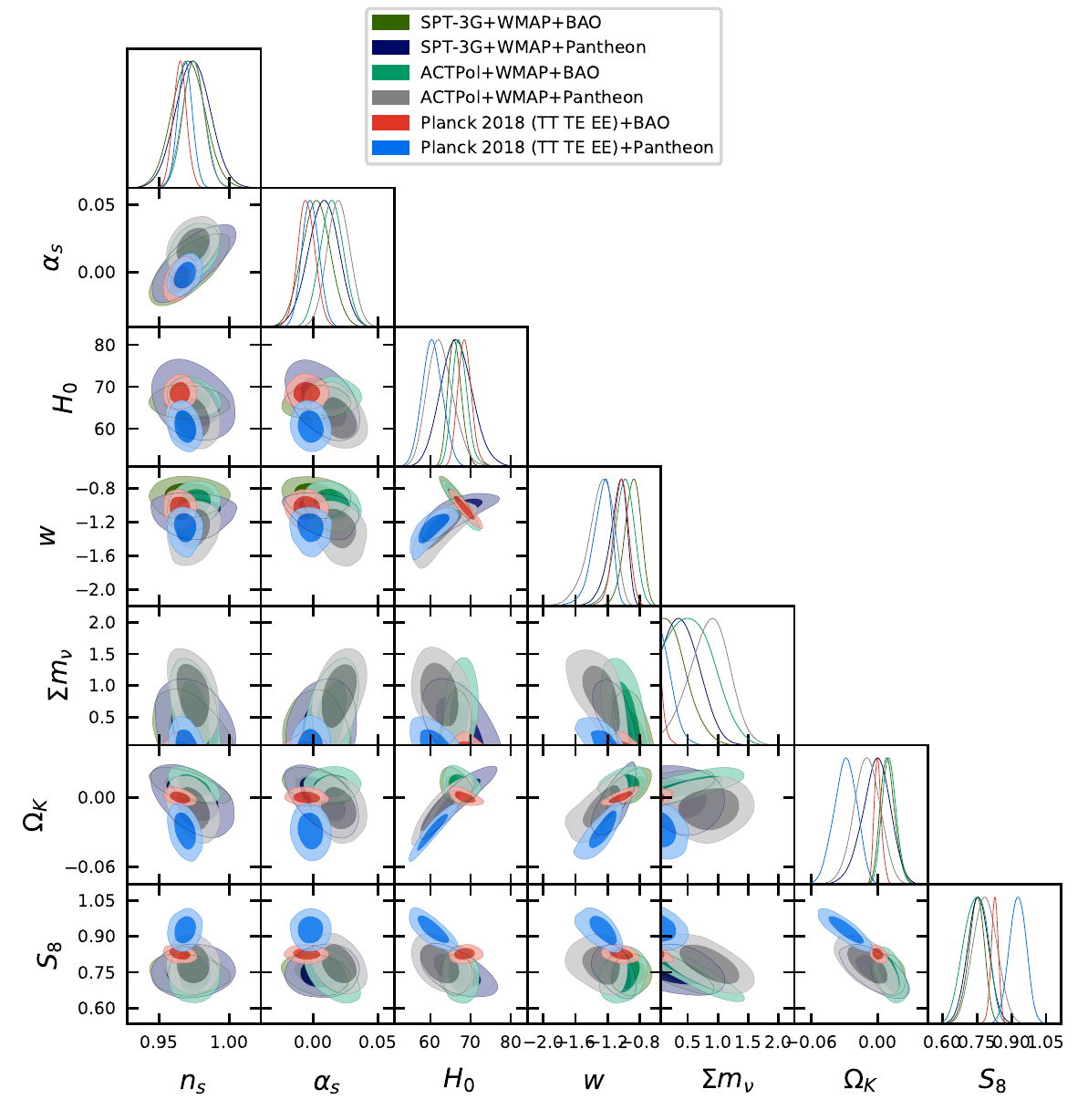}
\caption{Triangular plot showing the 1D posterior distributions and 2D contour plots for some of the parameters in $w$CDM$+ \Omega_k + \sum m_{\nu} + \alpha_s$, with the inclusion of BAO and Pantheon data.}
\label{fig:womegakmnunrunBP}
\end{figure}

Now, we reintroduce the curvature parameter in the sample, varying at the same time also the running of the scalar spectral index $\alpha_s\doteq dn_s/d\log k$ and the total mass of neutrinos. In this case we instead ignore the effective number of relativistic degrees of freedom and remain with 10 independent parameters. The numerical constraints can be found in \autoref{tab.wCDM+Omk+Mnu+nrun}, and we show the correlation plots in \autoref{fig:womegakmnunrun} and \autoref{fig:womegakmnunrunBP}. SPT-3G+WMAP alone has not enough constraining power to reach the convergence for this scenario, so we do not report this case in \autoref{tab.wCDM+Omk+Mnu+nrun} and \autoref{fig:womegakmnunrun}.

\begin{itemize}
\item[$H_0$)] Analyzing the Planck (TT TE EE) observations of the cosmic microwave background, we constrain $H_0=53.2^{+5.8}_{-16}$ Km/s/Mpc alleviating the tension with SH0ES within $2\sigma$ because of the the large error-bars, as we can see also in \autoref{fig:womegakmnunrun}. This shift is completely due to a volume effect, as shown by the forecasts. Including BAO and Pantheon we get $H_0=68.6^{+1.5}_{-1.8}$ Km/s/Mpc and $H_0=60.5\pm 2.5$ Km/s/Mpc, respectively. These values are in tension with SH0ES at $2.4\sigma$ and $4.6\sigma$. As concerns ACTPol+WMAP the result is in line with Planck even though slightly shifted towards smaller values ($H_0=50.9^{+5.3}_{-14}$ Km/s/Mpc). Also in this case, the tension is alleviated within $2\sigma$. The inclusion of BAO and Pantheon basically leads to the same conclusion already discussed for Planck. Finally, for SPT+WMAP+BAO and SPT+WMAP+Pantheon we obtain $H_0=66.1^{+1.5}_{-1.8}$ Km/s/Mpc and $H_0=66.5\pm3.9$ Km/s/Mpc, in line with the other datasets.
\item[$\Omega_k$)] In this extended framework, Planck (TT TE EE) indicates a closed Universe at slightly more than $2\sigma$ ($-0.162<\Omega_k<-0.006$ at 95\% CL). Combining Planck (TT TE EE)+BAO only improves the bounds on the curvature parameter, severely reducing the room allowed for deviations from flatness. Planck (TT TE EE)+Pantheon show instead a mild $2.6\sigma$ preference for a closed Universe ($\Omega_k=-0.0291^{+0.011}_{-0.0097}$). As concerns ACTPol+WMAP, this dataset is consistent with a flat spacetime geometry at slightly more than one standard deviation. Combining ACTPol+WMAP+BAO, as well as SPT-3G+WMAP+BAO, we find a shift towards positive $\Omega_k$, but flatness is still recovered within $1.5\sigma$. Finally, ACTPol+WMAP+Pantheon is in perfect agreement with $\Omega_k=0$ as well as SPT-3G+WAMP+Pantheon.

\item[$w$)] The dataset Planck (TT TE EE) weakly constrains the Dark Energy equation of state to $w=-1.55^{+1.0}_{-0.75}$. Including BAO we get the more constraining bound $w=-1.038^{+0.098}_{-0.088}$ that is in perfect agreement with the cosmological constant. On the other hand, Planck(TT TE EE)+Pantheon show a mild $2\sigma$ preference for phantom Dark Energy ($w=-1.27^{+0.14}_{-0.087}$). ACTPol+WMAP and its combination with BAO and Pantheon are both in line with Planck. In particular, from ACTPol+WMAP+Pantheon we still observe a slight preference for $w<-1$ at about $1.6\sigma$. As concerns SPT-3G+WMAP+BAO and SPT-3G+WMAP+Pantheon, they are both consistent with $w=-1$ within 1 standard deviation.

\item[$S_8$)] We confirm the Planck preference for a larger $S_8$ value also in this extended parameter space. In particular, Planck (TT TE EE) gives $S_8=0.989^{+0.095}_{-0.063}$ while including BAO and Pantheon we obtain  $S_8=0.826 \pm 0.016$ and $S_8=0.927 \pm 0.037$, respectively. On the other hand, ACTPol+WMAP constrains $S_8=0.872^{+0.092}_{-0.076}$ only partially supporting the Planck result. Anyway from ACTPol+WMAP+BAO and ACTPol+WMAP+Pantheon we recover lower values for $S_8\sim0.74-0.78$, that are also in line with SPT-3G+WMAP+BAO and SPT-3G+WMAP+Pantheon. Therefore we again observe discrepancies about the value of $S_8$ predicted by different experiments and CMB datasets, even if for a safe comparison we should consider the weak lensing data $S_8$ estimated in this extended model, that are likely to be extremely relaxed because of the volume effect.

\item[$M_{\nu}$)] As usual, analyzing the Planck (TT TE EE) data we do not find evidence for a neutrino mass, with the final limit reading $\sum m_{\nu}=0.43^{+0.16}_{-0.37}$ eV, attributable entirely to a volume effect, as suggested by the forecasts. This bound becomes much tighter including BAO ($\sum m_{\nu}<0.0799$ eV) while Planck (TT TE EE)+Pantheon gives $\sum m_{\nu}<0.209$ eV. In this model ACTPol+WMAP prefer a non-vanishing total neutrino mass at the high statistical level of $3.8\sigma$ ($\sum m_{\nu}=1.17\pm 0.31$ eV) suggesting values around the eV range. This preference is reduced to $2.9\sigma$ for ACTPol+WMAP+Pantheon ($\sum m_{\nu}=0.86^{+0.34}_{-0.30}$ eV), while considering ACTPol+WMAP+BAO we find a less significant $\sum m_{\nu}=0.61^{+0.22}_{-0.48}$ eV, consistent also with massless neutrinos at $1.3\sigma$. Finally, SPT-3G+WMAP+BAO give only an upper limit of $\sum m_{\nu}$, while SPT-3G+WMAP+Pantheon again an indication for massive neutrinos at slightly more than 1$\sigma$.

\item[$\alpha_s$)] Exploiting the Planck (TT TE EE) observations of the Cosmic Microwave Background, we do not find evidence for a running of the spectral index. Including BAO and Pantheon does not change the result: all these datasets are in perfect agreement with $\alpha_s=0$. Conversely, in this case ACTPol+WMAP prefer a positive running $0.0006<dn_{s}/d\log k<0.0456$ at more than 99\% CL (see also Ref.~\cite{Forconi:2021que}). This preference is reduced at about $2.2\sigma$ for ACTPol+WMAP+Pantheon ($n_{s}/d\log k =0.0195 \pm 0.0089$) and at $1.6\sigma$ for ACTPol+WMAP+BAO ($n_{s}/d\log k =0.0143 \pm 0.0090$). Finally, SPT-3G+WMAP+BAO and SPT-3G+WMAP+Pantheon are consistent with vanishing running, in line with Planck. 
\end{itemize}


\subsection{\boldmath{$w$}\textbf{CDM}  \boldmath{$+ \Omega_k + N_{\rm eff} + \alpha_s$}}
\label{sec.wCDM+Omk+Neff+nrun}

\begin{table*}[htbp!]
\begin{center}
\renewcommand{\arraystretch}{1.5}
\resizebox{1\textwidth}{!}{\begin{tabular}{c |c c c c|  c c c | c   c }
	\hline
	\textbf{Parameter} & \textbf{\nq{Planck\\ (LCDM forecasts)}}  & \textbf{\nq{Planck 2018\\ (TT TE EE)}}  &  \textbf{\nq{\\ +BAO}} & \textbf{\nq{\\ +Pantheon}} & \textbf{ACTPol+WMAP}  & \nq{\\\textbf{+BAO}} & \nq{\\\textbf{+Pantheon}}  & \nq{\textbf{SPT-3G+WMAP}\\ \textbf{+BAO}} & \nq{\textbf{SPT-3G+WMAP} \\ \textbf{+ Pantheon}} \\
	\hline\hline
	$\Omega_{\rm b} h^2$ &$0.02240 \pm 0.00032$&$0.02254\pm 0.00026$ &$0.02220 \pm 0.00025$ & $0.02250 \pm 0.00026$ & $0.02170 \pm 0.00045$&$0.02168 \pm 0.00040$ & $0.02175 \pm 0.00041$ & $0.02265\pm 0.00040$ &$0.02270 \pm 0.00041$\\
	
	$\Omega_{\rm c} h^2$ &$0.1208 \pm 0.0039$&$0.1172 \pm 0.0034$ &$0.1164 \pm 0.0033$ & $0.1174 \pm 0.0034$ & $0.1099^{+0.0052}_{-0.0061}$&$0.1098^{+0.0054}_ {-0.0060}$ & $0.1100^{+0.0052}_ {-0.0061}$ & $0.1157^{+0.0058}_ {-0.0067}$ &$0.1157^{+0.0062}_{ -0.0070}$ \\
	
	$100\,\theta_{\rm {MC}}$ &$1.04085 \pm 0.00051$&$1.04127 \pm 0.00049$ &$1.04136 \pm 0.00049$ & $1.04126 \pm  0.00048$ & $1.04289 \pm 0.00097$&$1.0430 \pm 0.0010$ & $1.04291 \pm 0.00098$ & $1.03963 \pm 0.00087$ &$1.03964 \pm 0.00088$\\
	
	$\tau$   &$0.052 \pm 0.011$& $0.0489 \pm 0.0085$ &$0.0560 \pm 0.0082$ & $0.0497 \pm 0.0086$ & $0.056 \pm 0.013$ & $0.058 \pm 0.013$ &$0.056 \pm 0.013$& $0.060 \pm 0.013$ &$0.059 \pm 0.013$\\
	
	$\log(10^{10}A_{\rm S})$ &$3.042 \pm 0.023$& $3.027 \pm 0.019$ &$3.040 \pm 0.019$ & $3.029 \pm 0.020$ & $3.021^{+0.035}_{-0.031}$&$3.025 \pm 0.032$ & $3.021 \pm 0.032$ & $3.041 \pm 0.030$ &$3.038 \pm 0.032$\\
	
	$n_s$ &$0.967 \pm 0.015$&$0.967\pm 0.012$ &$0.952 \pm 0.012$ & $0.965 \pm 0.012$ & $0.938 \pm 0.028$&$0.934 \pm 0.026$ & $0.940 \pm 0.027$ & $0.976 \pm 0.025$ &$0.980 \pm 0.026$ \\
	
	$w$ &$-0.87^{+0.59}_{-0.31}$&$-1.30^{+0.89}_{-0.47}$ & $-1.03^{+0.10}_{-0.092}$ & $\mathbf{-1.209^{+0.098}_{-0.075}[^{+0.16}_{-0.18}]}$ &$-0.81^{+0.73}_{-0.30}$&$-0.914^{+0.11}_{-0.093}$ & $-1.047^{+0.083}_{-0.064}$ & $\mathbf{-0.835^{+0.095}_{-0.086}[^{+0.17}_{-0.19}]}$ &$-0.971^{+0.071}_{-0.054}$ \\
	
	$\Omega_k$ &$-0.0195^{+0.032}_{-0.0082}$&$\mathbf{-0.042^{+0.036}_{-0.012}[^{+0.041}_{-0.031}]}$ &$0.0019^{+0.0031}
	_{-0.0042}$ & $\mathbf{-0.025^{+0.011}_{ -0.010}[^{+0.020}_{-0.021}]}$ & $-0.022^{+0.042}_{-0.013}$&$0.0096^{+0.0054}_{-0.0067}$ & $0.002^{+0.013}_{-0.011}$ & $0.0053^{+0.0055}_{-0.0070}$ &$0.002^{+0.013}_{-0.011}$ \\
	
	$N_{\rm eff}$ &$3.10 \pm 0.30$&$2.97 \pm 0.24$ & $2.78 \pm 0.23$ & $2.96 \pm 0.24$ &$2.31^{+0.42}_{-0.49}$&$2.26^{+0.39}_{-0.46}$ & $2.35^{+0.40}_{-0.48}$ & $3.07^{+0.42}_{-0.50}$ &$3.12^{+0.44}_{-0.52}$ \\
	
	$\alpha_s$ &$0.0008 \pm 0.0092$&$-0.0032 \pm 0.0081$ &$-0.0110 \pm 0.0083$ & $-0.0040\pm 0.0084$ & $-0.006 \pm 0.015$&$-0.008 \pm 0.014$ & $-0.005 \pm 0.015$ & $0.004 \pm 0.015$ &$0.007 \pm 0.016$ \\
	
	$H_0$ [Km/s/Mpc] &$62^{+9}_{-20}$&$61^{+9}_{-20}$ &$67.2 \pm 2.1$ & $61.3^{+2.3}_{-2.6}$ & $58^{+9}_{-20}$&$63.1^{+2.3}_{-2.6}$ & $65.1^{+3.2}_{-3.7}$ & $65.6\pm2.4$ &$70.0^{+3.9}_{-4.6}$ \\
	
	$\sigma_8$&$0.774^{+0.079}_{-0.17}$&$0.835^{+0.095}_{-0.19}$ &$0.810 \pm 0.027$ & $0.825 \pm 0.015$ & $0.737^{+0.082}_{-0.18}$&$0.768 \pm 0.033$ & $0.798 \pm 0.024$ & $0.746 \pm 0.032$ &$0.785 \pm 0.025$ \\
	
	$S_8$ &$0.880 \pm 0.088$&$0.950^{+0.095}_{-0.084}$ &$0.820 \pm 0.015$ & $0.923 \pm 0.039$ & $0.87^{+0.12}_{-0.10}$&$0.807 \pm 0.025$ & $0.816 \pm 0.049$  & $0.774 \pm 0.027$ &$0.766 \pm 0.047$ \\
	
	$\Omega_m$&$0.44^{+0.11}_{-0.28}$&$0.44^{+0.12}_{-0.28}$ &$0.308 \pm 0.016$ & $0.376 \pm 0.029$ & $0.50^{+0.43}_{-0.36}$&$0.332 \pm 0.019$ & $0.314^{+0.028}_{-0.032}$  & $0.324 \pm 0.017$ &$0.286^{+0.027}_{-0.030}$ \\
	
	$r_{\rm drag}$ [Mpc] &$146.6\pm2.8$&$148.1 \pm 2.3$ & $149.7 \pm 2.3$ & $148.1 \pm 2.3$ &$154.8 \pm 4.7$&$155.2 \pm 4.5$ & $154.6 \pm 4.5$  & $147.9 \pm 4.2$ &$147.7 \pm 4.3$ \\
	\hline \hline
	
	\end{tabular}}
\end{center}
\caption{Constraints on cosmological parameters at 68\% CL [95\% CL] for $w$CDM$+ \Omega_k + N_{\rm eff} + \alpha_s$.}
\label{tab.wCDM+Omk+Neff+nrun}
\end{table*}

\begin{figure}
\centering
\includegraphics[width=0.8\textwidth]{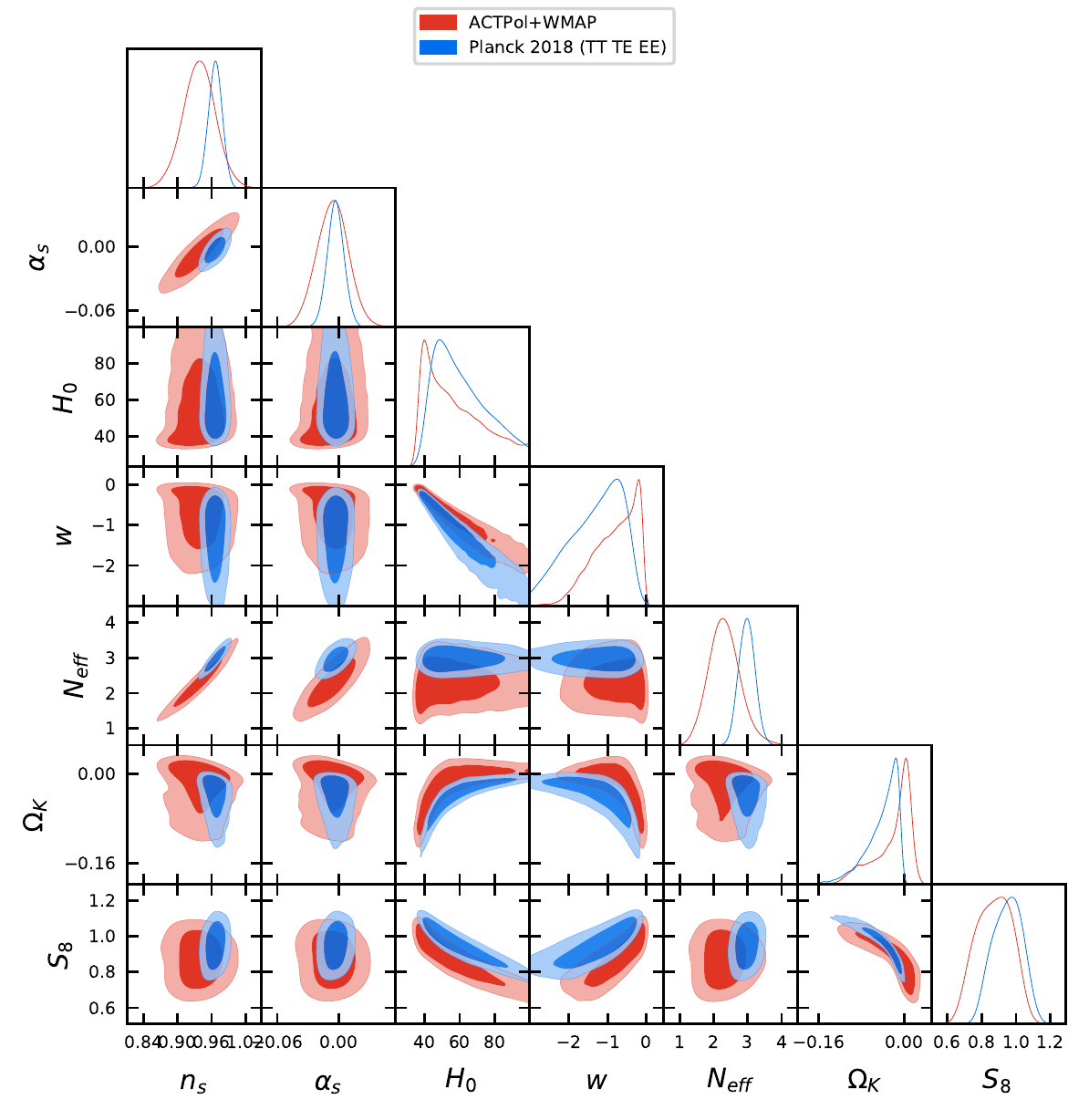}
\caption{Triangular plot showing the 1D posterior distributions and 2D contour plots for some of the parameters in $w$CDM$+ \Omega_k + N_{\rm eff} + \alpha_s$.}
\label{fig:womegaknnunrun}
\end{figure}

\begin{figure}
\centering
\includegraphics[width=0.85\textwidth]{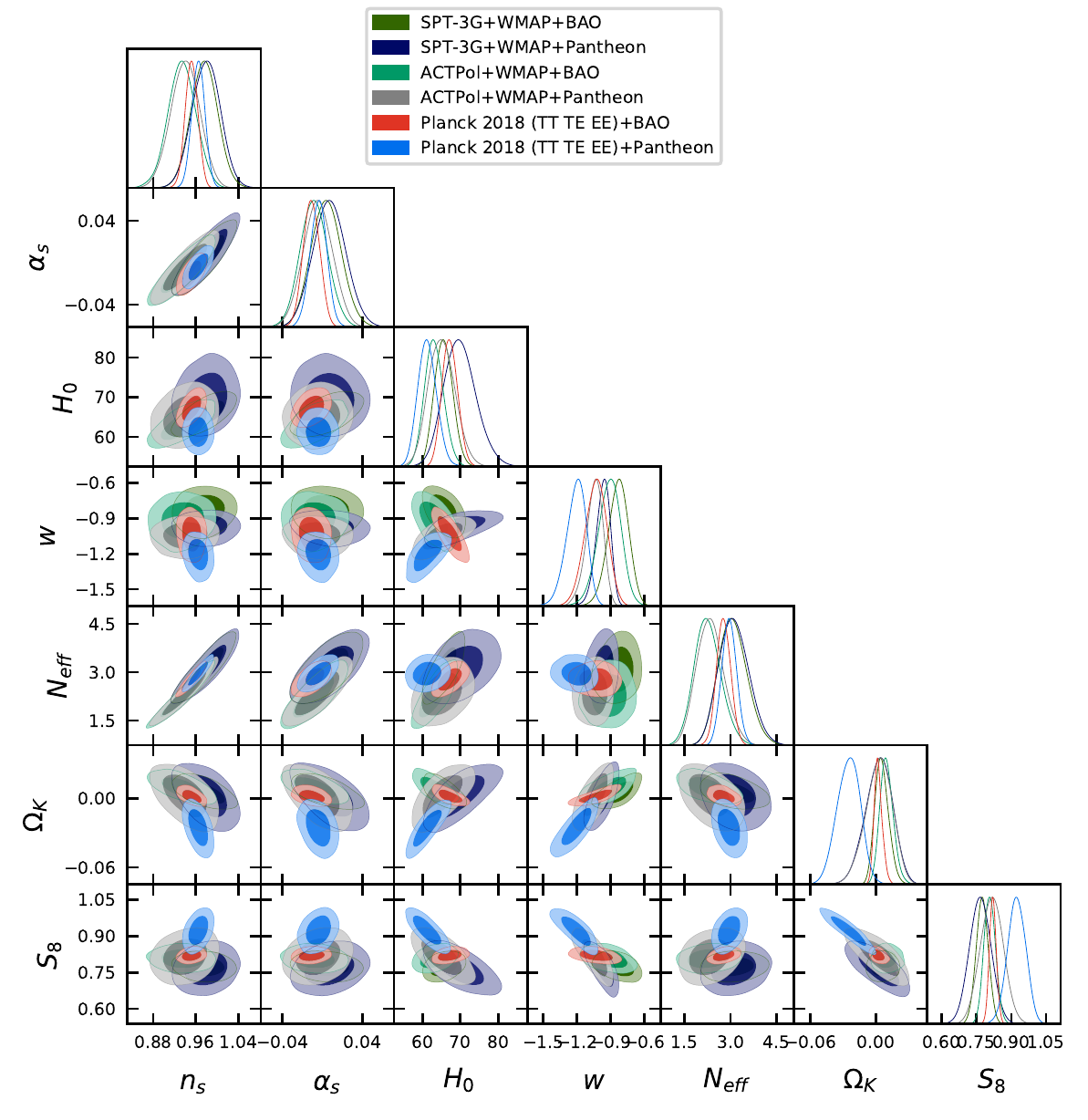}
\caption{Triangular plot showing the 1D posterior distributions and 2D contour plots for some of the parameters in $w$CDM$+ \Omega_k + N_{\rm eff} + \alpha_s$, with the inclusion of BAO and Pantheon data.}
\label{fig:womegaknnunrunBP}
\end{figure}

Here we replace the total mass of neutrinos with the effective number of relativistic neutrinos $N_{\rm eff}$, keeping always 10 independent free parameters in the model. The constraints on the parameters can be found in \autoref{tab.wCDM+Omk+Neff+nrun}, and we show the correlation plots in \autoref{fig:womegaknnunrun} and \autoref{fig:womegaknnunrunBP}. Also in this case, SPT-3G+WMAP alone has not enough constraining power to reach the convergence, so we do not report this case in \autoref{tab.wCDM+Omk+Neff+nrun} and \autoref{fig:womegaknnunrun}.

\begin{itemize}
\item[$H_0$)] The Hubble constant is poorly constrained to be $H_0=61^{+9}_{-20}$ Km/s/Mpc by Planck (TT TE EE) data, in line with our forecasts. Thus this dataset can be easily put into agreement with SH0ES in this extended parameter space, because of a volume effect. Instead, when BAO and Pantheon are included we get $H_0=67.2\pm2.1$ Km/s/Mpc ($2.5\sigma$ tension with SH0ES) and $H_0=61.3^{+2.3}_{-2.6}$ Km/s/Mpc, ($4.7\sigma$ tension with SH0ES), respectively. As concerns the Planck-independent observations of the cosmic microwave background, from ACTPol+WMAP we obtain $H_0=58^{+9}_{-20}$ Km/s/Mpc, \textit{i.e.}, weakly constrained and in line with Planck. Combining ACTPol+WMAP+BAO we get $H_0=63.1^{+2.3}_{-2.6}$ Km/s/Mpc and observe a preference for smaller values of the expansion rate with respect to Planck(TT TE EE)+BAO. This increases the  Hubble tension to $4\sigma$. Conversely, ACTPol+WMAP+Pantheon gives $H_0=65.1^{+3.2}_{-3.7}$ Km/s/Mpc reducing the tension with SH0ES to $2.4\sigma$. As concerns the South Pole Telescope data, from the combination SPT-3G+WMAP+BAO we obtain $H_0=65.6\pm2.4$ Km/s/Mpc ($2.8\sigma$ tension with SH0ES) while it is worth noting that SPT-3G+WMAP+Pantheon give $H_0=70.0^{+3.9}_{-4.6}$ Km/s/Mpc which is in perfect agreement with the direct local measurements.

\item[$\Omega_k$)] Planck (TT TE EE) prefers a closed Universe at slightly more than 95\% CL, \textit{i.e.}, $-0.106<\Omega_k-0.001$. Planck (TT TE EE)+BAO is instead in perfect agreement with flatness. Conversely, Planck (TT TE EE)+Pantheon shows a mild $2.3\sigma$ preference for a closed Universe. On the other hand, ACTPol+WMAP, SPT-3G+WMAP and their combinations with BAO and Pantheon are all in good agreement with $\Omega_k=0$ within two standard deviations.

\item[$w$)] Exploiting the  Planck (TT TE EE) data we obtain $w=-1.30^{+0.89}_{-0.47}$ and so, due to the large uncertainty, we can recover $w=-1$ within one standard deviation. Notice however that for the same reason we cannot rule out different behaviours. Combining Planck (TT TE EE)+BAO we find a perfect agreement with the baseline value. Conversely, Planck (TT TE EE)+Pantheon show a mild preference for a phantom dark energy ($w=-1.209^{+0.098}_{-0.075}$) at the level of $2.1\sigma$. ACTPol+WMAP and its combination with BAO and Pantheon are all consistent with the cosmological constant at one standard deviation. Instead, from SPT-3G+WMAP+BAO we find a $2\sigma$ preference for quintessential Dark Energy with the bounds reading $w=-0.835^{+0.095}_{-0.086}$. Finally, SPT-3G+WMAP+Pantheon is in good agreement with $w=-1$.

\item[$S_8$)] Also in this model, we observe discordant results for $S_8$. In particular, we confirm the Planck preference for $S_8\sim 0.9$ with the constraint for Planck (TT TE EE) reading $S_8= 0.950^{+0.095}_{-0.084}$. This recast a preference for larger $\sigma_8 > 0.8$ and $\Omega_{m}=0.44^{+0.12}_{-0.28}$ poorly constrained but shifted towards higher values, as well. Combining Planck (TT TE EE)+BAO we instead obtain $S_8=0.820\pm0.015$, closer to the $\Lambda$CDM value. However from Planck (TT TE EE)+Pantheon we recover the preference for both a larger matter density and a larger $\sigma_8$ that produces $S_8=0.923\pm0.039$. Exploiting ACTPol+WMAP data we infer $S_8=0.87^{+0.12}_{-0.10}$ that partially supports the Planck result but that is also consistent with cosmic shear measurements (obtained assuming a $\Lambda$CDM model and its minimal extensions~\cite{Heymans:2020gsg,KiDS:2020ghu,DES:2021vln,DES:2022ygi}) because of the large error-bars. Furthermore, for this dataset, including BAO and Pantheon we obtain always $S_8\sim 0.8$ that is given by a combination of $\sigma_8\sim 0.77-0.80$ and $\Omega_m\sim 0.3$. Finally, also SPT-3G+WMAP+BAO and SPT-3G+WMAP+Pantheon strongly prefer $S_8\sim0.77$ with the bounds reading $S_8=0.774\pm0.027$ and $S_8=0.766\pm0.047$, respectively. Consequently, we observe some mild discrepancies at the statistical level of $\sim 2\sigma$ about the results obtained by the different CMB experiments. 

\item[$N_{\rm{eff}}$)] As concerns the effective number of relativistic degrees of freedom, Planck (TT TE EE) and its combination BAO and Pantheon are all in good agreement with the reference value of three active neutrinos which is in fact recovered always within one standard deviation. Instead, ACTPol+WMAP gives $N_{\rm eff}= 2.31^{+0.42}_{-0.49}$ showing a preference for less radiation at the level of about $1.8\sigma$. The same preference can be observed for ACTPol+WMAP+BAO ($N_{\rm eff}= 2.26^{+0.39}_{-0.46}$) and ACTPol+WMAP+Pantheon ($N_{\rm eff}= 2.35^{+0.40}_{-0.48}$), respectively at the level of  $2\sigma$ and $1.7\sigma$. On the other hand, both SPT-3G+WMAP+BAO and SPT-3G+WMAP+Pantheon are in perfect agreement with the baseline predictions.

\item[$\alpha_s$)] In this case, we do not find evidence for a running of the spectral index and all the different datasets are consistent with $\alpha_s=0$ at most within $1.3\sigma$. 
\end{itemize}


\subsection{\boldmath{$w$}\textbf{CDM}  \boldmath{$+\Omega_k + \sum m_{\nu} + N_{\rm eff}  + \alpha_s$}}
\label{sec.wCDM+Omk+Mnu+Neff+nrun}

\begin{table*}[htbp!]
\begin{center}
\renewcommand{\arraystretch}{1.5}
\resizebox{1 \textwidth}{!}{\begin{tabular}{c | c c c c| c  c  c  |c c   }
	\hline
	\textbf{Parameter} & \textbf{\nq{Planck\\ (LCDM forecasts)}}  & \textbf{\nq{Planck 2018\\ (TT TE EE)}}  &  \textbf{\nq{\\ +BAO}} & \textbf{\nq{\\ +Pantheon}} & \textbf{ACTPol+WMAP}  & \nq{\\\textbf{+BAO}} & \nq{\\\textbf{+Pantheon}}  & \nq{\textbf{SPT-3G+WMAP}\\ \textbf{+BAO}} & \nq{\textbf{SPT-3G+WMAP}\\ \textbf{+ Pantheon}} \\
	\hline\hline
	
	$\Omega_{\rm b} h^2$ &$0.02229\pm 0.00033$&$0.02251\pm 0.00027$ & $0.02216 \pm 0.00025$ & $0.02247 \pm 0.00027$& $0.02170 \pm 0.00038$& $0.02147 \pm 0.00040$ &  $0.02160 \pm 0.00038$ & $0.02256\pm 0.00041$ &$0.02261 \pm 0.00041$ \\
	
	$\Omega_{\rm c} h^2$ &$0.1214 \pm 0.0039$&$0.1181 \pm 0.0034$ & $0.1165 \pm 0.0033$ & $0.1176 \pm 0.0033$& $0.1136 \pm 0.0057$& $0.1102 \pm 0.0055
	$ & $0.1115^{+0.0051}_{-0.0060}$ &  $0.1174^{+0.0062}_{-0.0069}$ &$0.1172^{+0.0061}_{-0.0070}$ \\
	
	$100\,\theta_{\rm {MC}}$ &$1.04053\pm 0.00055$&$1.04100\pm0.00050$ & $1.04133^{+0.00046}_{-0.00052}$ & $1.04116 \pm 0.00048$& $1.04212^{+0.00091}_{-0.0010}$&$1.04279 \pm 0.00098$ &$1.04251^{+0.00091}_{-0.0010}$ &  $1.03933 \pm 0.00090$ &$1.03928 \pm 0.00089$ \\
	
	$\tau$   &$0.051 \pm 0.010$&$0.0477 \pm 0.0081$ & $0.0557 \pm 0.0082$ & $0.0507 \pm 0.0080$& $0.054 \pm 0.013$ &$0.061 \pm 0.012$ & $0.059 \pm 0.012$ & $0.063 \pm 0.013$ &$0.061 \pm 0.013$\\
	
	$\log(10^{10}A_{\rm S})$ &$3.037 \pm 0.022$&$3.025 \pm 0.018$& $3.040 \pm 0.019$ & $3.031 \pm 0.019$ & $3.021 \pm 0.031$& $3.029 \pm 0.032$&$3.027 \pm 0.031$   & $3.050\pm 0.031$ &$3.043 \pm 0.031$\\
	
	$n_s$ &$0.966 \pm 0.015$& $0.968 \pm 0.012$ & $0.951\pm 0.012$ & $0.965 \pm 0.012$ & $0.950 \pm 0.025$&$0.928 \pm 0.025$ & $0.940 \pm 0.025$ &  $0.972 \pm 0.026$ &$0.980 \pm 0.026$ \\
	
	$\Omega_k$ &$-0.035^{+0.040}_{-0.014}$&$\mathbf{-0.076^{+0.060}_{-0.025}[^{+0.072}_{-0.096}]}$& $0.0026^{+0.0031}_{-0.0041}$ & $\mathbf{-0.029\pm 0.011[^{+0.021}_{-0.022}]}$ & $-0.046^{+0.047}_{-0.021}$&$\mathbf{0.0165 \pm 0.0069[^{+0.014}_{-0.015}]}$ &$-0.002^{+0.013}_{-0.012}$ &  $0.0090^{+0.0065}_{-0.0072}$ &$-0.002^{+0.014}_{-0.012}$\\
	
	$w$ &$-1.22^{+0.94}_{-0.42}$&$-1.51^{+0.96}_{-0.72}$& $-1.05^{+0.10}_{-0.091}$ & $\mathbf{-1.29^{+0.14}_{-0.093}[^{+0.23}_{-0.25}]}$ & $-1.32^{+1.1}_{-0.54}$&$-1.02^{+0.14}_{-0.11}$ &$-1.27^{+0.18}_{-0.11}$ &  $-0.895^{+0.12}_{-0.096}$ &$-1.08^{+0.13}_{-0.076}$\\
	
	$\sum m_{\nu}$ &$0.59_{-0.52}^{+0.23}$&$<0.553$ & $< 0.135$ & $< 0.239$& $\mathbf{1.01\pm0.34}$&$0.54^{+0.23}_{-0.29}[^{+0.69}_{-0.64}]$ & $\mathbf{0.72 \pm 0.28[^{+0.54}_{-0.55}]}$ & $< 0.427$ &$0.46^{+0.10}_{-0.40}$\\
	
	$N_{\rm eff}$ &$3.13\pm0.31$&$3.03 \pm 0.24$& $2.77 \pm 0.23$ & $2.97 \pm 0.24$& $2.63 \pm 0.45$&$\mathbf{2.21^{+0.36}_{-0.43}[^{+0.82}_{-0.73}]}$ & $2.41^{+0.38}_{-0.46}$ &  $3.10^{+0.43}_{-0.50}$ &$3.18^{+0.43}_{-0.53}$ \\
	
	$d n_s / d\log k$ &$0.005\pm0.10$&$-0.0004 \pm 0.0084$ & $-0.0111 \pm 0.0082$ & $-0.0035 \pm 0.0083$& $0.012 \pm 0.015$& $-0.006^{+0.012}_{-0.014}$& $0.003 \pm 0.014$ & $0.003 \pm 0.015$ &$0.010 \pm 0.016$\\
	
	$H_0$ [Km/s/Mpc] &$59^{+7}_{-20}$&$52^{+6}_{-20}$ & $67.4^{+1.9}_{-2.2}$ & $60.0 ^{+2.3}_{-2.7}$& $51.3^{+4.7}_{-14}$&$63.8^{+2.4}_{-2.7}$ &$61.3^{+2.9}_{-3.5}$  & $66.4 \pm 2.6$ &$67.0^{+3.9}_{-4.6}$\\
	
	$\sigma_8$&$0.714^{+0.071}_{-0.16}$&$0.731^{+0.075}_{-0.15}$ & $0.802 \pm 0.028$ & $0.802^{+0.028}_{-0.019}$& $0.621^{+0.052}_{-0.12}$&$0.699\pm 0.045$ &$0.695^{+0.038}_{-0.045}$ &  $0.715^{+0.043}_{-0.038}$ &$0.729^{+0.049}_{ -0.039}$\\
	
	$S_8$ &$0.877^{+0.091}_{-0.082}$&$0.990^{+0.094}_{-0.068}$ & $0.813^{+0.017}_{-0.015}$ & $0.922 \pm 0.039$ & $0.861^{+0.094}_{-0.077}$&$0.741 \pm 0.041$ &$0.778 \pm 0.050$ &  $0.746^{+0.039}_{-0.032}$ &$0.756 \pm 0.048$ \\
	
	$\Omega_m$&$0.52^{+0.15}_{-0.32}$&$0.62^{+0.21}_{-0.34}$ & $0.309 \pm 0.016$ & $0.398^{+0.032}_{-0.036}$& $0.64^{+0.25}_{-0.29}$&$0.338 \pm 0.020$ &$0.378 \pm 0.041$  & $0.327 \pm 0.018$ &$0.325^{+0.035}_{-0.044}$\\
	
	$r_{\rm drag}$ [Mpc] &$146.2 \pm 2.8$&$147.4 \pm 2.3$ & $149.8 \pm 2.3$ & $148.0\pm 2.3$& $151.2 \pm 4.4$&$155.3 \pm 4.3$ &$153.5 \pm 4.3$ &  $147.4 \pm 4.3$ &$146.9 \pm 4.3$ \\
	
	\hline \hline
	
	\end{tabular}}
\end{center}
\caption{Constraints on cosmological parameters at 68\% CL [95\% CL] for $w$CDM$+ \Omega_k + \sum m_{\nu} + N_{\rm eff} + \alpha_s$.}
\label{tab.wCDM+Omk+Mnu+Neff+nrun}
\end{table*}

\begin{figure}
\centering
\includegraphics[width=0.85\textwidth]{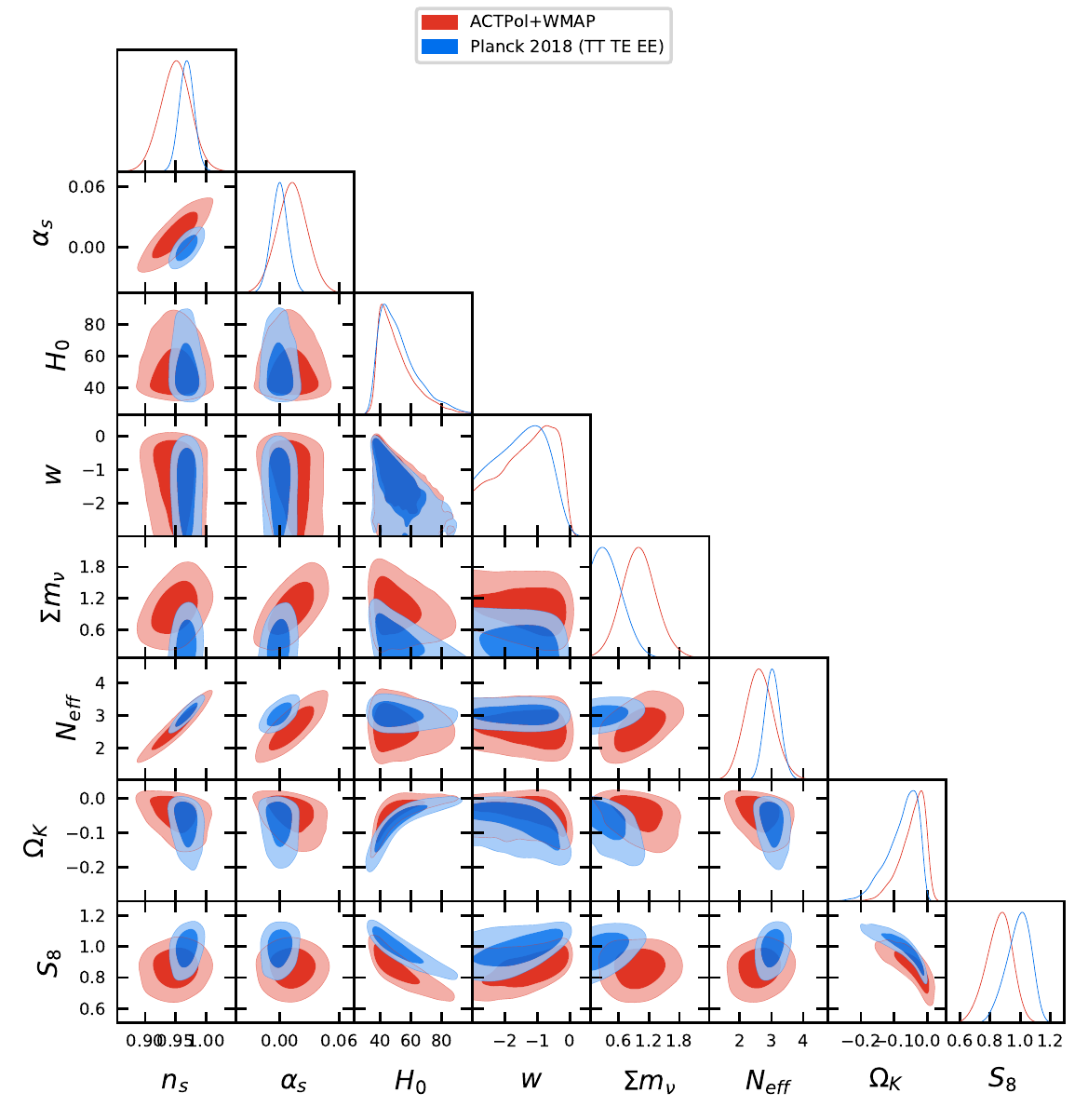}
\caption{Triangular plot showing the 1D posterior distributions and 2D contour plots for some of the parameters in $w$CDM$+ \Omega_k + \sum m_{\nu} + N_{\rm eff} + \alpha_s$.}
\label{fig:womegakmnunnunrun}
\end{figure}

\begin{figure}
\centering
\includegraphics[width=0.85\textwidth]{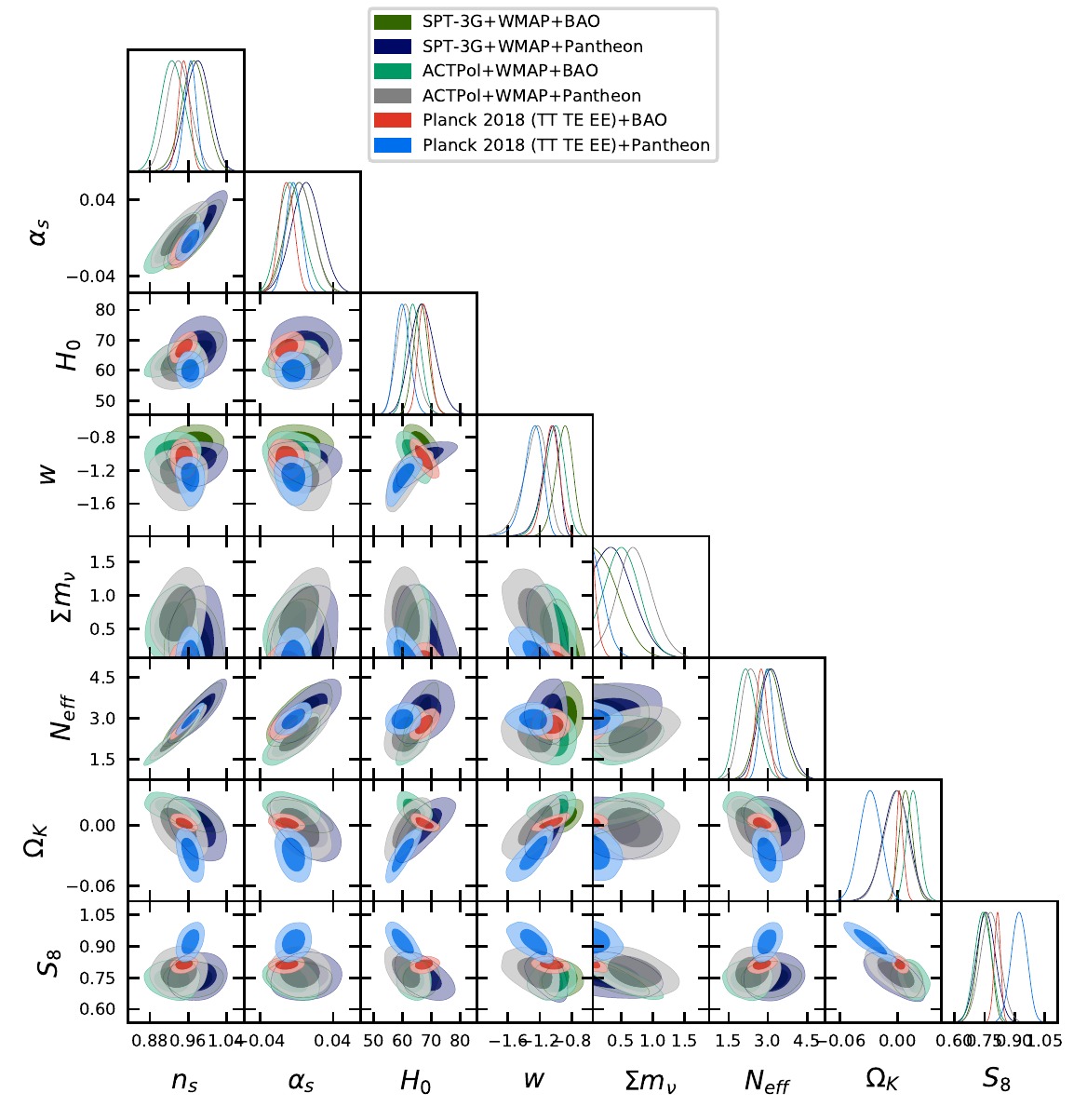}
\caption{Triangular plot showing the 1D posterior distributions and 2D contour plots for some of the parameters in $w$CDM$+ \Omega_k + \sum m_{\nu} + N_{\rm eff} + \alpha_s$, with the inclusion of BAO and Pantheon data.}
\label{fig:womegakmnunnunrunBP}
\end{figure}

This is the last extended cosmological model analyzed in this work. Here, we simultaneously vary the dark energy equation of state, the curvature parameter, the total mass of neutrinos, the effective number of relativistic degrees of freedom and the running of the scalar spectral index. Therefore we have a total number of 11 independent parameters. The results obtained from our MCMC analysis are summarized in \autoref{tab.wCDM+Omk+Mnu+Neff+nrun}, and we show the correlation plots in \autoref{fig:womegakmnunnunrun} and \autoref{fig:womegakmnunnunrunBP}. We remind the reader that also in this case, SPT-3G+WMAP alone has not enough constraining power to reach the convergence, so we do not report this case in \autoref{tab.wCDM+Omk+Mnu+Neff+nrun} and \autoref{fig:womegakmnunnunrun}.

\begin{itemize}

\item[$H_0$)] In this model with 11 degrees of freedom, the Planck (TT TE EE) data constrain the Hubble constant to $H_0=52^{+6}_{-20}$ Km/s/Mpc, consistent with the forecasted value, and because of the large errors, with SH0ES within $2\sigma$. Combining Planck (TT TE EE)+BAO we gain some constraining power and we get $H_0=67.4^{+1.9}_{-2.2}$ Km/s/Mpc, close to the baseline result of $\Lambda$CDM but in smaller tension with SH0ES ($2.6\sigma$) because of the bigger uncertainty due to the volume effect. Conversely, from Planck (TT TE EE)+Pantheon we obtain a preference for smaller expansion rate $H_0=60.0^{+2.3}_{-2.7}$ Km/s/Mpc. Consequently, the tension is increased at the level of $4.6\sigma$. Analyzing ACTPol+WAMP we obtain $H_0=51.3^{+4.7}_{-14}$ Km/s/Mpc. Also in this case, because of the large error-bars, the Hubble constant produced is in agreement within $2\sigma$ with SH0ES. Notice that the preference for small $H_0$ observed from this dataset remains robust also including BAO and Pantheon since we get $H_0=63.8^{+2.4}_{-2.7}$ Km/s/Mpc and $H_0=61.3^{+2.9}_{-3.5}$ Km/s/Mpc, respectively. Both the results are in tension with direct measurement at more than $3.5\sigma$. Instead, combining SPT-3G+WMAP+BAO and SPT-3G+WMAP+Pantheon we partially recover familiar-$\Lambda$CDM values of $H_0$, with the results reading $H_0=66.4\pm2.6$ Km/s/Mpc and $H_0=67.0^{+3.9}_{-4.6}$ Km/s/Mpc, respectively. In this case the $H_0$-tension is reduced to $2.4\sigma$ for SPT-3G+WMAP+BAO while SPT-3G+WMAP+Pantheon turns out to be in agreement with SH0ES at $1.5\sigma$. This is of course an effect due to the increase of the volume of the parameter space.

\item[$\Omega_k$)] As concerns the spacetime geometry, Planck (TT TE EE) prefers a closed Universe at more than $2\sigma$ ($-0.172<\Omega_k<-0.004$ at 95\% CL) while Planck (TT TE EE)+BAO well within one standard deviation ($\Omega_k=0.0026^{+0.0031}_{-0.0041}$). The combination Planck (TT TE EE)+Pantheon shows instead a preference for a closed Universe giving $\Omega_k<0$ at more than 99\% CL ($-0.059<\Omega_k<-0.001$). Analyzing the Planck independent CMB data, from the combination ACTPol+WMAP we find $\Omega_k=-0.046^{+0.047}_{-0.021}$, consistent with flatness within $1\sigma$. However, including BAO we get $\Omega_k>0$ at $2.4\sigma$ ($\Omega_k=0.0165\pm0.0069$) while ACTPol+WMAP+Pantheon is in agreement with flatness. SPT-3G+WMAP+BAO partially confirm the preference for a positive $\Omega_k$ already observed for ACTPol+WMAP+BAO, but it is reduced to $1.4\sigma$ ($\Omega_k=0.0090^{+0.0065}_{-0.0072}$). Finally, SPT-3G+WMAP+Pantheon are in good agreement with $\Omega_k=0$.

\item[$w$)] The Dark Energy equation of state is constrained to be $w=-1.51^{+0.96}_{-0.72}$ by Planck (TT TE EE) and it is consistent with $w=-1$ well within one standard deviation. Similarly, from Planck(TT TE EE)+BAO we get $w=-1.05^{+0.10}_{-0.091}$. Instead from Planck(TT TE EE)+Pantheon we observe the same preference for Phantom Dark Energy already discussed in other models with the constraint reading $w=-1.29^{+0.25}_{-0.45}$ at more than 99\% CL. The results inferred for ACTPol are in line with Planck: $w$ is poorly constrained by ACTPol+WMAP ($w=-1.32^{+1.1}_{-0.54}$) and due to the large uncertainty it remains consistent with the baseline-$\Lambda$CDM case. Adding also BAO we get a more tight bound $w=-1.02^{+0.14}_{-0.11}$ consistent with the  cosmological constant. Combining ACTPol+WMAP+Pantheon we find $w<-1$ at $1.5\sigma$ ($w=-1.27^{+0.18}_{-0.11}$). Finally, both SPT-3G+WMAP+BAO and SPT-3G+WMAP+Pantheon are in perfect agreement with $w=-1$.

\item[$S_8$)] Also in this 11-Dimensional parameter space, we observe the same differences about the value of $S_8$ discussed so far. In particular Planck(TT TE EE) gives $S_8=0.990^{+0.094}_{-0.068}$ thus preferring significant higher values of the matter density parameter $\Omega_m$ ($\Omega_{m}=0.62^{+0.21}_{-0.34}$), but not of $\sigma_8$ that is instead lower. Combining Planck(TT TE EE)+BAO, we instead recover the familiar value $\Omega_m\sim 0.3$ for the matter budget of the Universe that results into a constraint $S_8=0.813^{+0.017}_{-0.015}$ consistent with the $\Lambda$CDM estimation, and higher than the Weak Lensing result (assuming $\Lambda$CDM and its minimal extensions~\cite{Heymans:2020gsg,KiDS:2020ghu,DES:2021vln,DES:2022ygi}). Instead from the Planck(TT TE EE)+Pantheon data combination we get $S_8=0.922\pm0.039$ basically reflecting the same preference for a higher matter component mentioned above. As concerns the Planck independent data, in this case ACTPol+WMAP give a middle constrain $S_8=0.861^{+0.095}_{-0.077}$ balancing a preference for higher values of $\Omega_m=0.64^{+0.25}_{-0.29}$ and smaller values of $\sigma_8=0.621^{+0.052}_{-0.12}$. As usual, including BAO we recover $\Omega_m\sim 0.3$ and the final result for ACTPol+WMAP+BAO is $S_8=0.741\pm0.041$. This is the same behaviour observed also for ACTPol+WMAP+Pantheon ($S_8=0.778\pm 0.050$), SPT-3G+WMAP+BAO ($S_8=0.746^{+0.039}_{-0.032}$) and  SPT-3G+WMAP+Pantheon ($S_8=0.756\pm0.048$), all of them closer to the Weak Lensing estimate assuming a $\Lambda$CDM model and its minimal extensions~\cite{Heymans:2020gsg,KiDS:2020ghu,DES:2021vln,DES:2022ygi}. As a result, when Pantheon is combined with the CMB experiments the value of $S_8$ inferred by Planck(TT TE EE)+Pantheon becomes in tension with  ACTPol+WMAP+Pantheon at $2.3\sigma$ and with SPT-3G+WMAP+Pantheon at $2.7\sigma$.

\item[$M_{\nu}$)] As usual, analyzing the Planck (TT TE EE) data we do not find evidence for a neutrino mass ($\sum m_{\nu}<0.55$ eV). Including BAO and Pantheon leads only to more constraining upper limits: $\sum m_{\nu}<0.135$ eV and $\sum m_{\nu}<0.239$ eV, respectively. The situation is very different for ACTPol+WMAP ($\sum m_{\nu}=1.01\pm0.34$ eV) that also in this case suggest massive neutrinos at $2.8\sigma$ and prefer relatively large mass values, in disagreement with Planck. This preference is reduced to $1.9\sigma$ including BAO and the posterior distribution function is in fact shifted towards smaller mass values ($\sum m_{\nu}=0.54^{+0.23}_{-0.29}$ eV). Conversely, from ACTPol+WMAP+Pantheon we obtain $\sum m_{\nu}=0.72\pm0.28$ eV, suggesting again a non-vanishing mass at $2.6\sigma$. Finally, it is worth noting that the ACTPol preference for large massive neutrinos is not supported by SPT-3G+WMAP+BAO and SPT-3G+WMAP+Pantheon. 

\item[$N_{\rm{eff}}$)] The Planck (TT TE EE) data, as well as their combination with BAO and Pantheon, are in good agreement with the Standard Model prediction for three active neutrinos. Similarly, analyzing ACTPol+WMAP we do not find any relevant deviation with respect to the baseline value. On the other hand, combining ACTPol+WMAP+BAO we find  $N_{\rm eff}=2.21^{+0.36}_{-0.42}$ and hence a mild $2.3\sigma$ preference for a smaller number of relativistic degrees of freedom in the early Universe. However this preference is reduced to $1.7\sigma$ for ACTPol+WMAP+Pantheon while it is not supported by SPT-3G+WMAP+BAO and SPT-3G+WMAP+Pantheon since they are both consistent with $N_{\rm eff}=3.044$ well within one standard deviation.

\item[$\alpha_s$)] We do not find evidence for a running of the spectral index and all the datasets are in agreement with $d n_s / d\log k=0$ at most within $1.2\sigma$.
\end{itemize}

\section{Discussion and Conclusions}
\label{sec.conclusion}

In this work, we have presented an updated data-analysis of the most recent observations of the Cosmic Microwave Background temperature anisotropies and polarization angular power spectra released by three independent experiments: the Planck satellite; the Atacama Cosmology Telescope, and the South Pole Telescope. The measurements provided by the last two Collaborations have been used in combination with the WMAP satellite 9-years observation data in order to reach an accuracy comparable with Planck, keeping at the same time these two datasets independent from it. Aimed to test the consistency of the results obtained by the three experiments in extended parameter-spaces, we have analyzed 8 cosmological models that differ from the baseline $\Lambda$CDM case by the inclusion of different combinations of additional degrees of freedom, with the aim of finding a viable
minimal extended model that can bring all the CMB experiments in agreement. For each model, we have performed a full Monte Carlo Markov Chain analysis deriving the observational constraints on the cosmological parameters for all the different combinations of data. In addition, in order to evaluate possible bias due to the large volume of the parameter-spaces, we have simulated Planck temperature anisotropies and polarization angular spectra assuming a best-fit flat $\Lambda$CDM model and realized a MCMC analysis also over the simulated data. Finally, in light of the increasing tensions between cosmological and astrophysical observations, we have tested the robustness of our findings by quantifying the impact on the results deriving from the inclusion of CMB-independent probes; namely the Baryon Acoustic Oscillation measurements and the Type Ia Supernovae distance moduli measurements from the Pantheon sample.\footnote{Adding such astrophysical datasets is particularly useful also to break the degeneracy among cosmological parameters. In this regard it is worth noting that, in most cases, the degeneracies we have are geometrical and due to the fact that the parameters produce a similar angular distance to recombination and epoch of equality. Moreover, changing the number of parameters that vary, changes also the direction of the correlation. This also depends on the angular scale sampled by each experiment. See Refs.~\cite{Efstathiou:1998xx,Melchiorri:2000px,Corasaniti:2007rf,Elgaroy:2007bv} for further details.} In \autoref{sec.results}, we have carried out a detailed analysis, discussing parameter by parameter the results inferred in each cosmological model. This systematic investigation of extended cosmologies led us to recognize and confirm some well-known anomalies already observed in the Planck data (such as the preference for a closed Universe of this dataset) and to consolidate the other Planck-independent anomalies already discussed for the Atacama Cosmology Telescope data (such as the indication a for smaller effective number of relativistic particles and massive neutrinos). In particular, analyzing the same extended models in light of different CMB experiments, we have found that such anomalies remain robust predictions regardless from the number of free fitting parameters. This suggests that they are not an artifact of the specific model, but rather an actual preferences of the data. Conversely, no significant anomalies are observed analyzing the South Pole Telescope data (although often this dataset do not have enough constraining power to produce constraints on the cosmological parameters in extended models). In what follows, we summarize our general conclusions for the different parameters and experiments.

\begin{itemize}
\item[$H_0$)] All the values of the expansion rate of the Universe inferred by the different observations of the Cosmic Microwave Background turn out to be largely sensitive to the underlying cosmological model and its assumptions. In extended cosmologies, $H_0$ is often poorly constrained by the CMB data and, due to the large error-bars, the Hubble tension is alleviated at the level of 2$\sigma$ (\textit{i.e.}, 95\% CL) in the vast majority of the models analyzed in this work, see also \autoref{fig:RC_H0}. This partially supports the statement that the current tension could be a manifestation of the inadequacy of the standard $\Lambda$CDM model of cosmology to correctly describe more precise observations from widely different epochs of the Universe and provides direct evidence that assuming this model introduces a bias in the constraints inferred by data. However, when the different CMB observations are combined with astrophysical datasets the errors are typically reduced and thus the tension increased. 
As many have previously speculated, we cannot exclude possible observational systematics in the local measurements of $H_0$, even if the new extensive analysis performed by the SH0ES collaboration~\cite{Riess:2021jrx}, with $\sim70$ different tests over all the possible systematics proposals in the literature, seems to suggest that this possibility is more and more unlikely. The current the $5.3\sigma$ tension between the latest local result~\cite{Riess:2022mme} and Planck suggests that an "unknown systematic error" speculation to account for the disagreement is not enough. Moreover, even excluding the SH0ES measurement and combining the alternative local measurements the $H_0$ tension still ranges in between $4-6.5\sigma$ (see Refs.~\cite{Verde:2019ivm,Riess:2019qba,DiValentino:2020vnx,DiValentino:2022fjm}). Finally, a 2-rung measurement without the SNIa still prefer a higher Hubble constant~\cite{Kenworthy:2022jdh}. In conclusion, while our analysis does not reveal the nature of the $H_0$-tension, it reinforces both the need for more precise independent observations of the Cosmic Microwave Background from future surveys and the importance of interpreting current and future observations in light of possible new physics beyond $\Lambda$CDM. 

\begin{figure}[htbp!]
\centering
\includegraphics[width=0.85\textwidth]{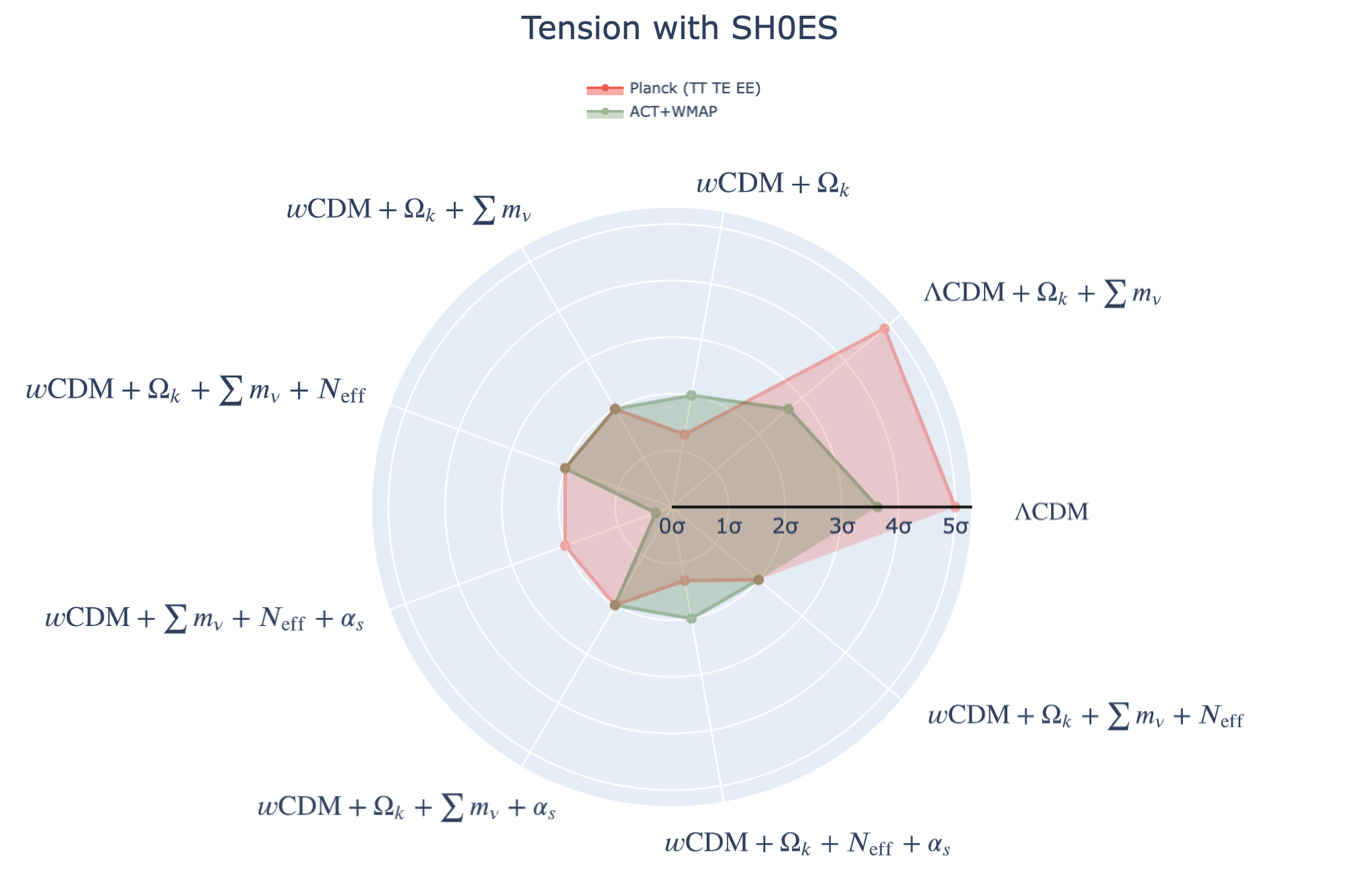}
\caption{Tension between the value of the Hubble parameter $H_0=73\pm 1$ measured by the SH0ES collaboration and the results inferred in different extended cosmological models by analyzing the Planck 2018 (red points) and ACTPol+WMAP (green points) observations of the Cosmic microwave Background angular power spectra.
}
\label{fig:RC_H0}
\end{figure}

\item[$\Omega_k$)] The Planck preference for a closed Universe related to the low amplitude of quadrupole and octupole modes observed in the Planck (TT TE EE) data (which remains essentially unexplained within the baseline flat $\Lambda$CDM model), but mainly due to the excess of lensing in the damping tail, is non significantly reduced in extended cosmologies, see also \autoref{fig:RC_Omk}. However, this dataset becomes more consistent with the inflationary prediction for a spatially flat Universe at about 2 standard deviations with the increase of the number of parameters, but this is not because of a real shift of the mean value, but for the increase of the errors. In addition, both Atacama Cosmology Telescope and South Pole Telescope data, once combined with WMAP, remain consistent with spatial flatness in extended parameter-spaces, as well. The results are basically stable under the combination of CMB and astrophysical observations, although sometimes we observe shifts from $\Omega_k=0$ (both towards positive and negative values) that never cross the level of $\sim 2.5 \sigma$. The tension on the curvature of the universe between CMB data hints for a possible undetected systematic error.
\begin{figure}[h!]
\centering
\includegraphics[width=0.85\textwidth]{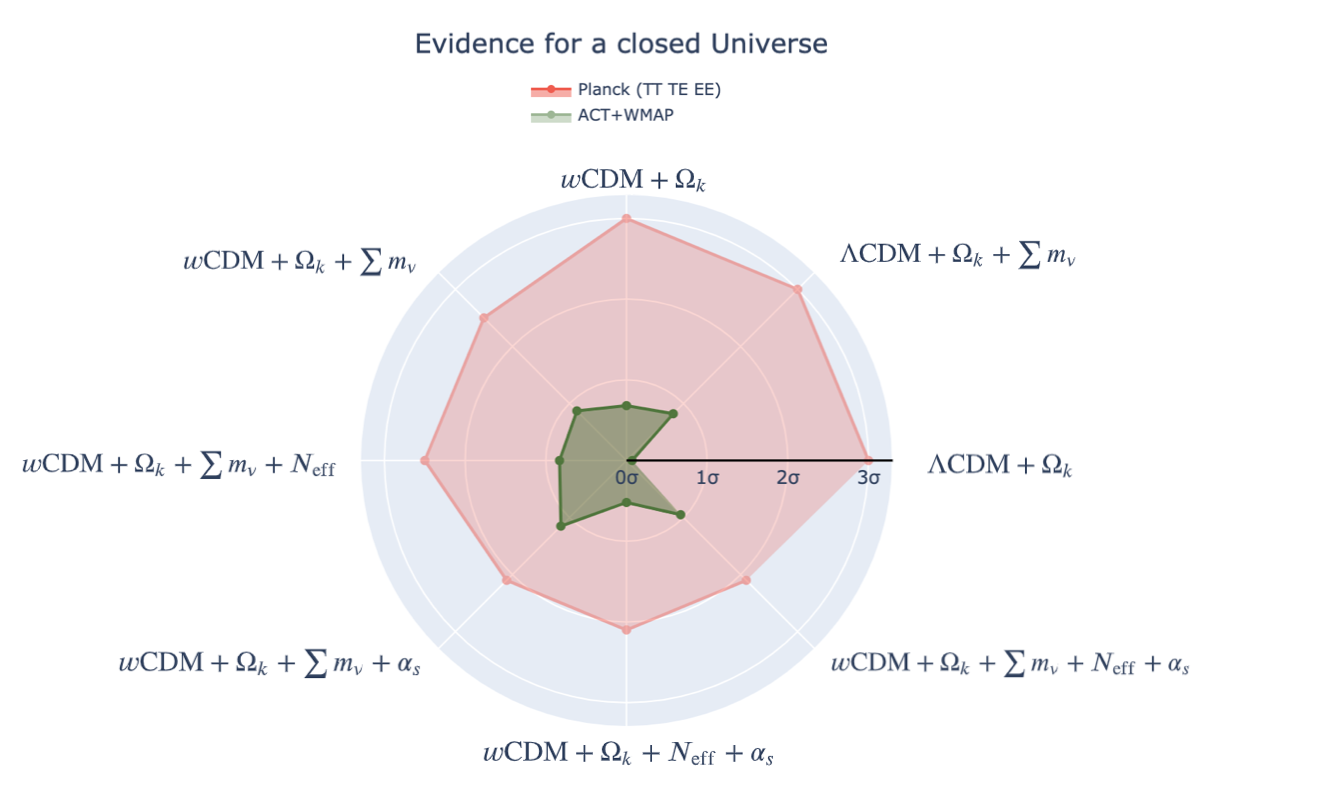}
\caption{Evidence for a closed Universe ($\Omega_k<0$) obtained analyzing the Planck 2018 (red points) and ACTPol+WMAP (green points) observations of the cosmic microwave background in extended cosmological models. }
\label{fig:RC_Omk}
\end{figure}

\clearpage
\item[$S_8$)] We observe some tensions about the value of the parameter $S_8$ inferred by the different experiments and extended models. In particular, analyzing the Planck (TT TE EE) measurements of the CMB temperature and polarization angular power spectra, we report a systematic preference for $S_8\gtrsim 0.9$, in disagreement with cosmic shear surveys results~\cite{Heymans:2020gsg,KiDS:2020ghu,DES:2021vln,DES:2022ygi}. This preference is only partially supported by the Atacama Cosmology Telescope and South pole Telescope data that, for many models, suggest instead $S_8\sim 0.7 - 0.8$, in line with cosmic shear measurements. In any case, it is worth noting that different values of $S_8$ often recast discordant results for the parameter $\sigma_8$ and the matter budget of the Universe, $\Omega_m$. In particular, the latter is very badly constrained in extended cosmologies and we can observe a shift towards higher values from all the CMB data. This shift is usually compensated by a preference for smaller $\sigma_8$ in ACT and SPT, but not in Planck. Including BAO and Pantheon measurements, we instead recover familiar values $\Omega_m\sim 0.3$ and thus smaller $S_8$. As a result, the constraints on $S_8$ and $\sigma_8$ obtained by the different experiments and extended models show inconsistencies at the level of $\sim 2\sigma-3\sigma$.

\item[$w$)] The different CMB experiments poorly constrain the Dark Energy Equation of state in extended parameter-spaces. So, because of the large error-bars, the results are typically consistent with a cosmological constant term in Einstein's equations within one standard deviation. However, for the same reason, the bounds are not enough constraining to rule out a different behaviour and so both phantom and quintessential models of Dark Energy remain consistent with observations. Combining the CMB data with BAO measurements the constraints usually shrink around $w=-1$ and we do not observe significant deviations from this baseline value. On the other hand, in \autoref{fig:RC_w} we show the results obtained considering the the Type Ia Supernovae distance moduli measurements from the Pantheon sample in combination with the CMB data. In this case from Planck and ACT we systematically observe a preference for phantom Dark Energy ($w<-1$) at a statistical level ranging between $1.5\sigma$ and $2.5\sigma$ while the combination of SPT and Pantheon data is usually consistent with the cosmological constant value at one standard deviation. 
\begin{figure}[htbp!]
\centering
\includegraphics[width=0.6\textwidth]{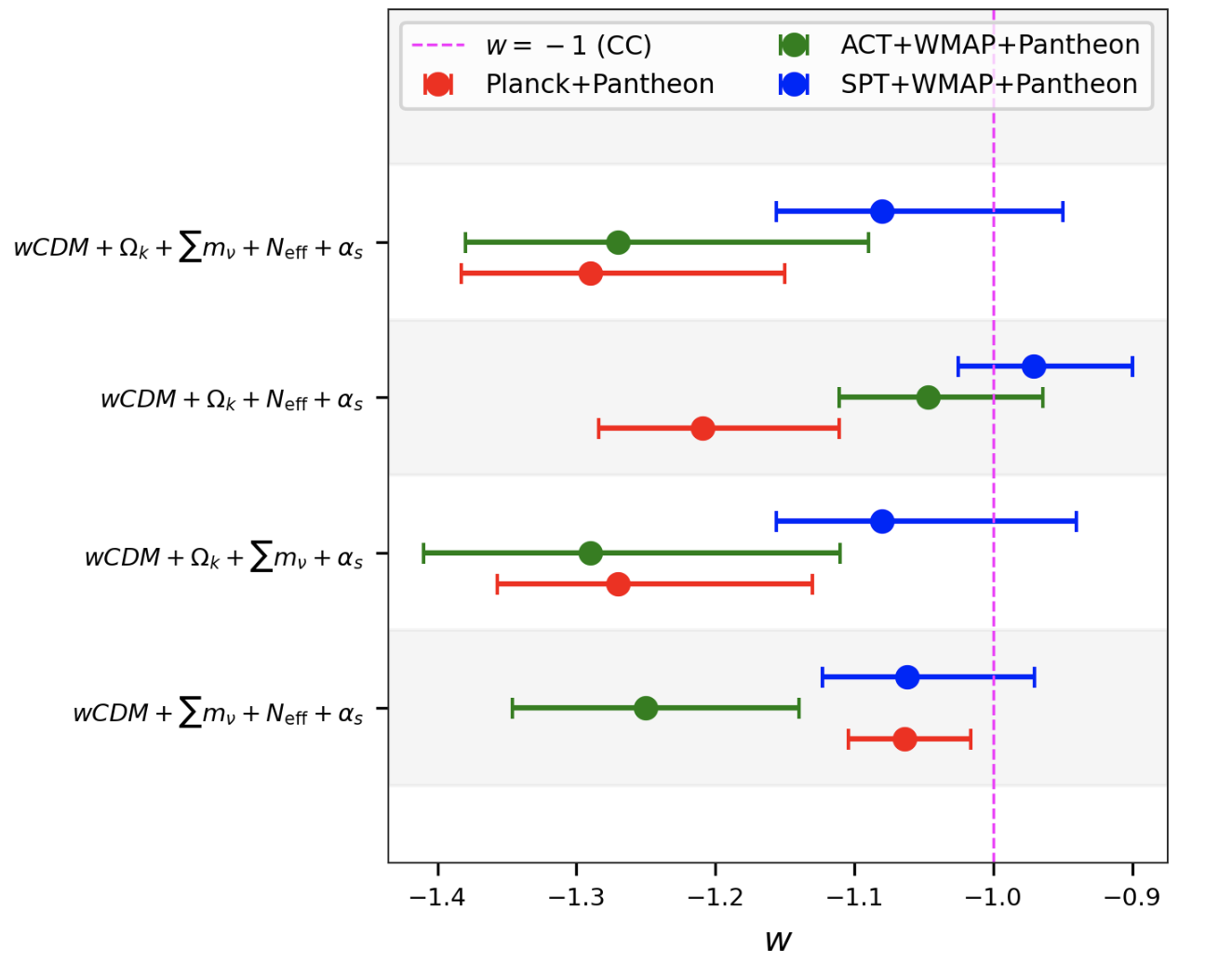}
\caption{Values inferred for the Dark Energy equation of state $w$ analyzing different observations of the cosmic microwave background in combination with the SNeIa distance moduli measurements from the Pantheon sample.}
\label{fig:RC_w}
\end{figure}

\item[$M_{\nu}$)] Planck and South Pole Telescope+WMAP data are always consistent with massless neutrinos within about one standard deviation. In general, both these datasets prefer smaller neutrino masses $\sum m_{\nu}\lesssim 0.5$ eV and including astrophysical observations only lead to obtain more constraining upper limits. Instead the situation is very different for the Atacama Cosmology telescope: as one can see in \autoref{fig:RC_Mnu}, when this dataset is combined with WMAP it always shows a moderate-to-strong preference ($2.5\sigma - 4\sigma$) for massive neutrinos, suggesting larger mass values $\sum m_{\nu}\gtrsim 0.5$ eV, in tension with Planck and SPT. When ACT is combined with BAO and Pantheon, this evidence, although slightly reduced, can be still observed, producing an interesting robust indication for massive neutrinos.
\begin{figure}[htbp!]
\centering
\includegraphics[width=0.9\textwidth]{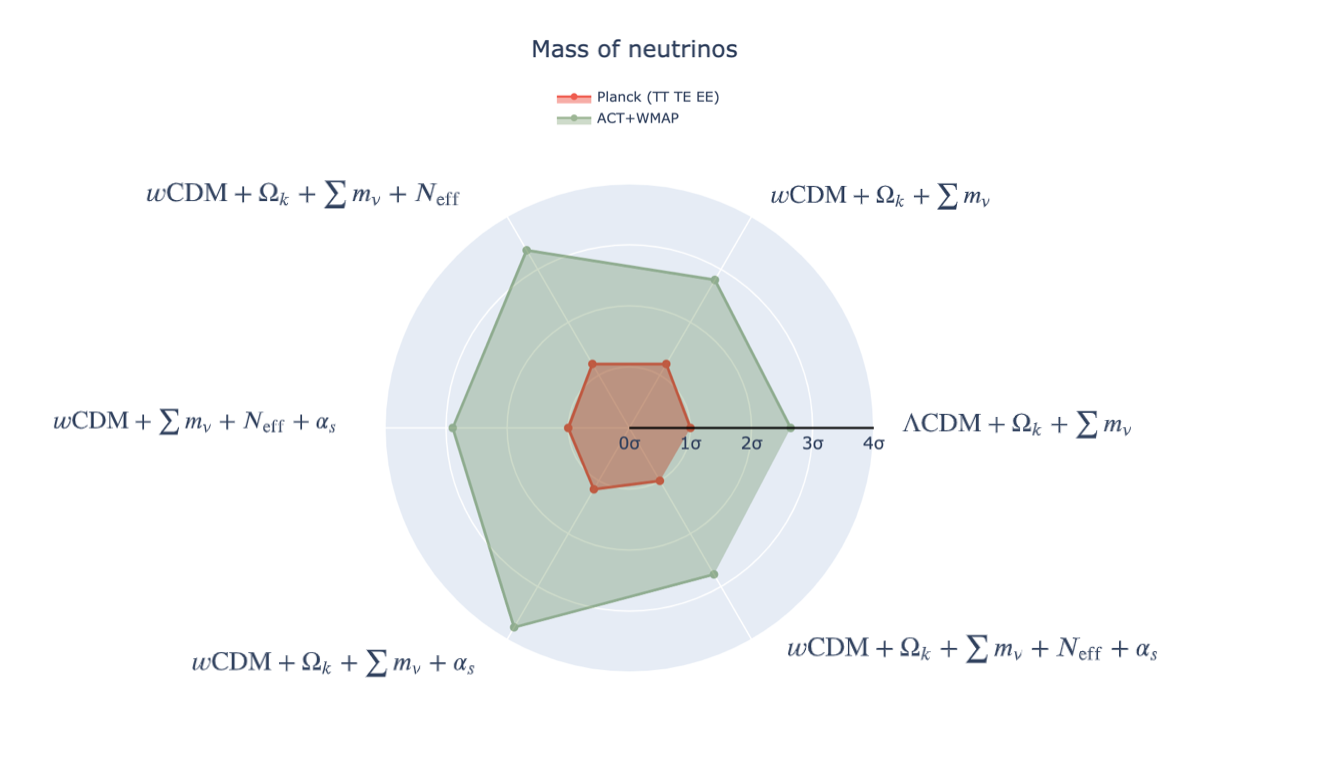}
\caption{Indication for a total neutrino mass obtained analyzing the Planck 2018 (red points) and ACTPol+WMAP (green points) observations of the cosmic microwave background in extended cosmological models.}
\label{fig:RC_Mnu}
\end{figure}

\item[$N_{\rm eff}$)] As concerns the effective number of relativistic degrees of freedom, we summarize our most constraining results for the different CMB experiments in \autoref{fig:RC_Neff}. In particular, we find that Planck and South Pole telescope (as well as their combination with BAO and Pantheon) are in good agreement with the value predicted by Standard Model for three active neutrinos ($N_{\rm eff}=3.044$) which is in fact recovered always within one standard deviation. Conversely, the Atacama Cosmology Telescope data show a preference for a smaller amount of radiation in the early Universe, with a statistical significance that changes between $1.8\sigma$ and $3\sigma$, depending on the cosmological model. This indication does not change significantly including astrophysical observations (BAO and Pantheon), but it becomes weaker increasing the number of free parameters. However this is not due to a shift of the mean value, but rather to the larger error-bars, see also \autoref{fig:RC_Neff}. 
\vspace{0.5cm}
\begin{figure}[htbp!]
\centering
\includegraphics[width=0.6\textwidth]{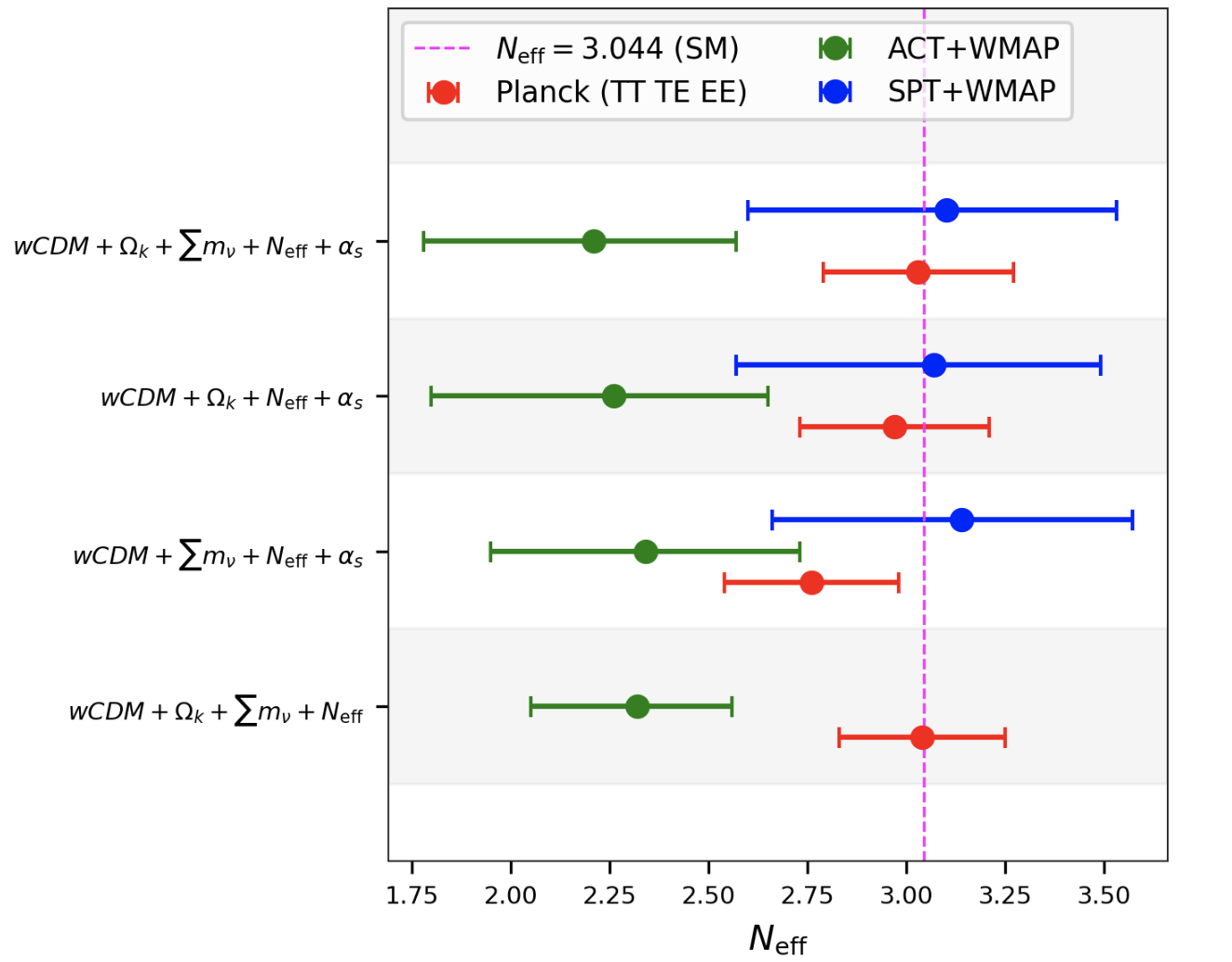}
\caption{Values inferred for the effective number of relativistic species ($N_{\rm eff}$) analyzing different observations of the cosmic microwave background.}
\label{fig:RC_Neff}
\end{figure}

\item[$\alpha_s$)] We summarize the results for the running of the spectral index of inflationary scalar modes in \autoref{fig:RC_nrun}.  Analyzing the Planck data and the South Pole Telescope data, we find no evidence for a running of the spectral index in any cosmological model. On the other hand, the results obtained for the Atacama Cosmology Telescope are more uncertain: the preference for a small positive running of this dataset is confirmed only in a few extended parameter-spaces. For instance, we find $d n_s/d\log k>0$ at $\sim 3\sigma$ within $w$CDM+$\Omega_k$+$\sum m_{\nu}$+$\alpha_s$, but in all the other models analyzed in this work the results remain consistent with  $d n_s/d\log k=0$ within $1\sigma$. Our analysis therefore suggests that the ACT preference for a positive running may be sensitive to the underlying degrees of freedom of the model and so that the same effects can be easily recast in different cosmological parameters.  

\begin{figure}[h!]
\centering
\includegraphics[width=0.6\textwidth]{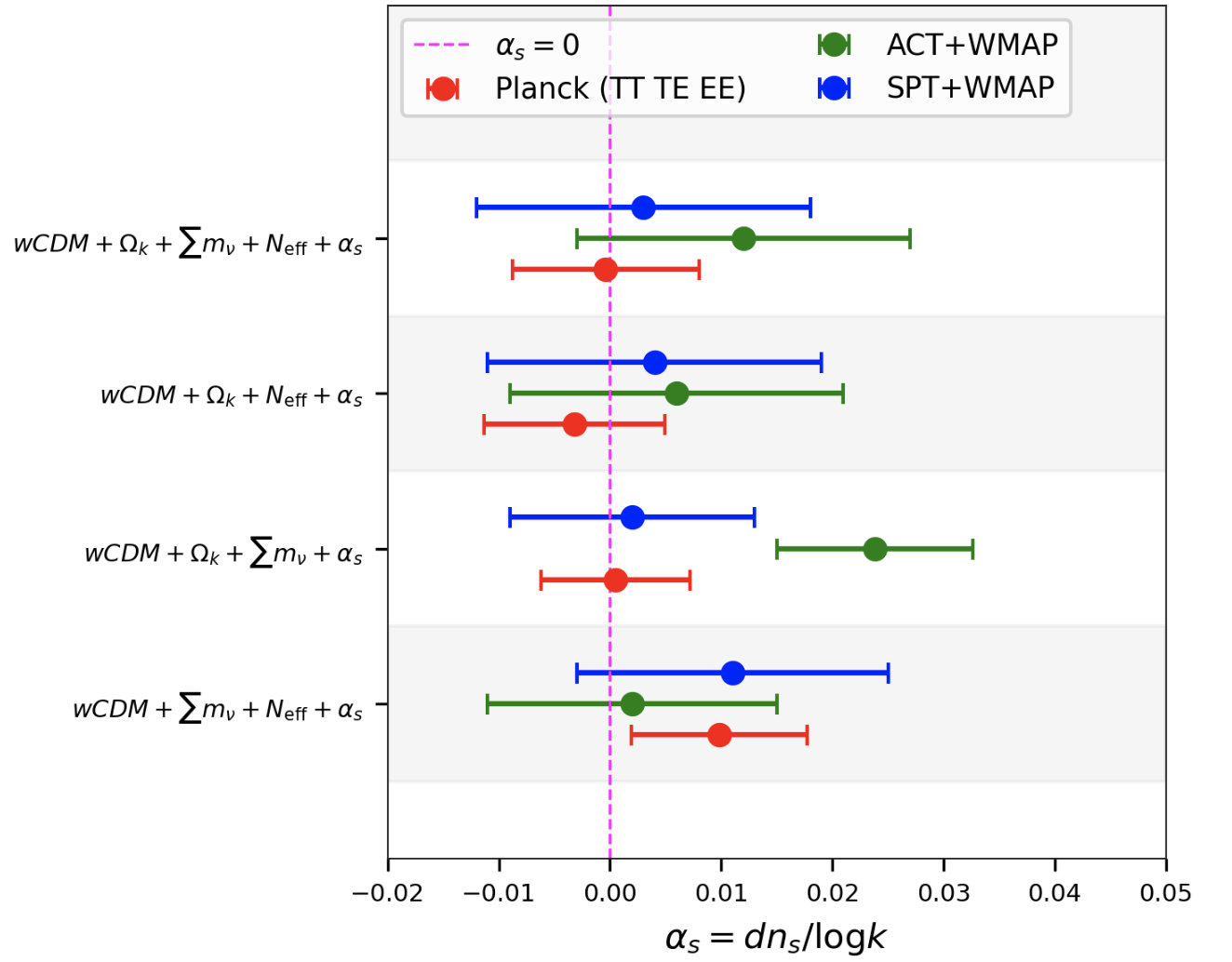}
\caption{Values of the running of the spectral index of inflationary scalar modes ($\alpha_S$) inferred analyzing different observations of the cosmic microwave background.}
\label{fig:RC_nrun}
\end{figure}
\end{itemize}
\newpage 

We conclude underlying that our analysis provides several convincing pieces of evidence for anomalies in the CMB angular power spectra that are hard to explain within the standard cosmological model. Once that some of the internal assumptions of the model have been relaxed, the results we obtain are not always consistent with what one would expect within the baseline case. An important point to keep in mind is that constraints obtained after marginalization over the nuisance but also "anomalous" parameters provide a more conservative and robust determination with respect to the results inferred with the extra parameters fixed at $\Lambda$CDM. On the other hand, given the different data combinations and especially the large number of analyzed cosmologies, one may ask whether our results could be interpreted in terms of a "look-elsewhere effect": a phenomenon in statistical data-analysis where an apparently statistically significant indication may actually arise by chance, for instance because of the sheer size of the parameter space. However, since the main tensions and anomalies highlighted in this work are observed also within minimal extensions to $\Lambda$CDM (where only one more parameter is added) and given that they remain robust in all the extended cosmologies and systematically for the same datasets, it seems highly unlikely that such self-consistent indications for diverging patterns could be a casual product of the statistics rather than an actual preference of the data.

In any case, we can basically outline two scenarios. On one side, we can trust the standard cosmological model and assume $\Lambda$CDM to be the correct paradigm of the Universe. In this case to explain our findings we need to conclude that significant unaccounted-for systematics in the data are producing biased results. Statistical fluctuations may in fact explain why independent measurements of the cosmic background angular power spectra (as well as their combination with astrophysical datasets) often point in the discordant (although self-consistent) directions and why the different anomalies are never supported by different experiments. However, within $\Lambda$CDM all the CMB experiments agree pretty well about the value of the expansion rate and the tension with local measurements remains basically unexplained. In this regard it is worth stressing one more time that the value of the $H_0$ inferred by observations of the cosmic microwave background is largely sensitive to the underlying cosmological model and for this reason the tension may easily be a product of the model rather than of the data. Therefore the second possibility we have is taking the data at the face value, leaving the study of observational systematics to the different collaborations that in many cases already explored this possibility deeply. 
In this case our analysis leads weight to the hypothesis that a solution of the current tensions may involve new physics beyond $\Lambda$CDM. Since we do not fully understand the physics underlying three major unknown ingredients of the standard cosmological model (\textit{i.e.}, Inflation, Cold Dark Matter and Dark Energy), this possibility deserves to be considered seriously, as well. In this regard, it is also worth noting that none of the different combinations of extra parameters is able to fully reconcile all the discordances. Therefore a satisfactory solution might require a more radical shift in the paradigm of cosmology, see for instance Refs~\cite{Smith:2022hwi,Poulin:2021bjr,Perivolaropoulos:2021jda,Dainotti:2021pqg,Renzi:2021xii, Dainotti2022, Marra:2021fvf, Camarena:2021jlr, DiValentino:2020zio, Jedamzik:2020zmd,CANTATA:2021ktz,DiValentino:2021izs,Schoneberg:2021qvd,Abdalla:2022yfr,DiValentino:2022fjm,Krishnan:2020vaf,Colgain:2022nlb,Colgain:2022rxy,Krishnan:2020obg,Colgain:2021beg} and the discussion therein. So, while our analysis is not conclusive about the nature of these tensions and anomalies and all the possibilities remain viable, it definitively reveals several discrepancies in the most recent CMB data that remain basically unexplained. Clearly, only future independent high-precision CMB and large scale structure measurements could provide a definitive answer~\cite{CMB-S4:2016ple,Abazajian:2019eic,CMB-S4:2022ght,SimonsObservatory:2018koc,SimonsObservatory:2019qwx,CLASS,LBIRD,CORE:2016ymi,CORE:2017oje,NASAPICO:2019thw,DESI:2013agm,DESI:2016fyo,EUCLID:2011zbd,EuclidTheoryWorkingGroup:2012gxx,Eifler:2020vvg,Eifler:2020hoy,CMB-HD:2022bsz,Sehgal:2020yja,Kollmeier:2019ogo,Chluba:2019kpb,Rhodes:2019lur,DiValentino:2020vhf,Chang:2022tzj,Blum:2022dxi}.


\begin{acknowledgments}
EDV is supported by a Royal Society Dorothy Hodgkin Research Fellowship. WG and AM are supported by "Theoretical Astroparticle Physics" (TAsP), iniziativa specifica INFN.
\end{acknowledgments}

\bibliography{bibfiles.bib}

\vfill
\end{document}